\documentclass[galaxies,review,accept,pdftex,oneauthor]{mdpi}
\firstpage{1} 
\pubvolume{1}
\issuenum{1}
\articlenumber{0}
\pubyear{2024}
\copyrightyear{2024}
\datereceived{ } 
\daterevised{ } % Comment out if no revised date
\dateaccepted{ } 
\datepublished{ } 
\hreflink{https://doi.org/} % If needed use \linebreak
\Title{A primer on the formation and evolution of hydrogen deficient\\ Central Stars of Planetary Nebulæ and Related Objects. }

\TitleCitation{A primer on the formation and evolution of hydrogen deficient Central Stars of Planetary Nebulæ and Related Objects.}

\Author{Marcelo M. Miller Bertolami $^{1,2}$\orcidA{}}

\AuthorNames{Marcelo M. Miller Bertolami}

\AuthorCitation{Miller Bertolami, M. M.}
\address{%
$^{1}$ \quad Instituto de Astrofísica de La Plata, Consejo Nacional de Investigaciones Científicas y Técnicas Avenida Centenario (Paseo del
  Bosque) S/N, B1900FWA La Plata; mmiller@fcaglp.unlp.edu.ar\\  
$^{2}$ \quad Facultad de Ciencias Astronómicas y Geofísicas, Universidad Nacional de La Plata Avenida Centenario (Paseo del Bosque) S/N, B1900FWA La Plata, Argentina.}

\corres{Correspondence: mmiller@fcaglp.unlp.edu.ar}

\abstract{ We present a brief review on the formation and evolution of
  hydrogen deficient central stars of planetary nebulae.  We  include a detailed description of the main observable features of both the central stars and their
  surrounding nebulae and review their main classifications. We also provide a brief description
  of the possible progenitor systems of hydrogen deficient central stars,
  as well as, of transients closely connected to the formation of these stars. 
  In particular we offer  a detailed theoretical explanation of  the
  main evolutionary scenarios, both single and binary, devised to explain these stars and nebulae. Particular emphasis is
  made in the description of the so-called born again scenario, their quantitative predictions and uncertainties.
Finally we discuss the pros and cons of both binary and single evolution channels, draw some conclusions and discuss open questions in the field. }


\begin{document}

%%%%%%%%%%%%%%%%%%%%%%%%%%%%%%%%%%%%%%%%%%
%\setcounter{section}{-1} %% Remove this when starting to work on the template.

\section{ Introduction: Planetary Nebulae and their central stars}\label{intro}
\label{sec:intro}
Contrary to widespread belief, the evolution of low-mass stars is far
from fully understood. This is especially true with regard to the
formation and evolution of Planetary nebulae (PNe) and their central
stars (CSPNe).  Discovered in 1764 by Messier
\citep{1781cote.rept..227M}, our current understanding of PNe was
initially drafted by the middle of the twentieth century when it was
suggested \cite{1956AZh....33..315S} that CSPNe could be the immediate
progenitors of white dwarf (WD) stars, and the descendants of red
giants\cite{1966PASP...78..232A}.  This scenario received a strong
theoretical foundation  from numerical stellar evolution
models \cite{1970AcA....20...47P,1971AcA....21..417P}. These models
showed that, if the envelope of giant stars is removed by very strong
winds, the bare nuclei of red giants evolve through of the regions of
the Hertzsprung-Russell diagram (HR) corresponding to the CSPNe, and
in the appropriate timescales (i.e. tens of thousands of
years). Later, several authors
\cite{1978ApJ...219L.125K,1985MNRAS.212..837K} developed a concrete
mechanism involving the interaction between the slow and dense wind of
the  asymptotic giant branch (AGB) star and the fast and tenuous wind of the CSPNe that explained
the global characteristics of PNe. These theoretical studies
solidified the idea that CSPNe and their PNe are transition objects
between the AGB and the WD cooling sequence.

By the beginning of the
twenty-first century the diversity of PNe morphologies together with
the discovery of of close binaries in the nuclei of some PNe suggested
that binarity might also be involved in the formation and/or shaping
of PNe
\citep{2000ASPC..199..115B,2002ARA&A..40..439B,2009PASP..121..316D,2017NatAs...1E.117J,2019ibfe.book.....B}. Searches
for companions through photometric monitoring and spectroscopic studies
have been carried out for decades. Bond \cite{2000ASPC..199..115B}
reported that 10\% to 15\% of a sample of about a hundred CSPNe had
companions with periods of less than 3 days, which corresponds to what
is expected in systems that went through a common envelope
phase. Subsequent studies showed an absence of systems with longer
periods \citep{2009A&A...505..249M,2009A&A...496..813M}.  Using data
from the Kepler space telescope, a fraction of close binaries in PNe
of 23.5\% was derived, with periods between 2 hours and 30 days
(almost all of them below 5 days
\cite{2021MNRAS.506.5223J}). Photometric searches for cold companions
using infrared excesses \cite{2015MNRAS.448.3132D} determined a
fraction of possible companions of $32\pm 16$\% or $52\pm 24$\% using
the I and J bands respectively. This determination includes distant
companions that will never interact with the primary. The detection and
confirmation of long period binaries is intrinsically complicated due
to the large periods involved, which is why they have only recently
started to be discovered. There are, currently, only 5 known PNe with
measured long period binaries, with periods between 140 and 5000 days
and eccentricities between 0 and 0.5\footnote{A regularly updated
  database is available on David Jones' website
  \url{https://www.drdjones.net/bcspn/} }
\cite{2019ibfe.book.....B}. In addition, there are about 12 confirmed
binary CSPNe with separations between 90 and 2500 AU. Triple CSPNe are
rare but possible.  To date only one of these systems (NGC~246) has
been confirmed as a triple system \citep{2014MNRAS.444.3459A} while
two others (LoTr~5 and Sp~3) have been proposed but not confirmed.  Even rarer, observations with the James Webb Space Telescope (JWST) of the PN NGC~3132 suggest that this system might habour at least a stellar quartet \cite{2022NatAs...6.1421D} . A
separate comment  is warranted for  the central stars where both components are
compact objects (double degenerate systems) in a very close binary,
which have been suggested as possible Type Ia SN progenitors. To date
only few systems have been studied in detail and all of them have
total masses below the Chandrasekhar mass \cite{2010AJ....140..319H,
  2010ApJ...714..178T, 2020A&A...638A..93R, 2022MNRAS.511.2033H}.

The formation of PNe not only requires material ejection
mechanisms able to account for the morphological variety observed but, above all, it requires  the
synchronization of the time scale associated with the expansion and
dissipation of the ejected material, and the time scale associated with
the contraction and heating of the central star (CS) that provides the UV
photons \citep[e.g.][]{2013A&A...558A..78J}. A direct consequence of
this is that not all stars that eject their envelope (either by winds
or binary interaction), and contract to form white dwarfs, will form
planetary nebulae.

The enormous diversity, not only of morphologies of the PNe but of
chemical compositions of both PNe and their CSs suggests that many
different evolutionary paths seem to be able to form PNe. More
detailed descriptions of the topics covered here and on the various
uses and interests of PNe for other areas of the astrophysics can be
found in reviews \cite{2022PASP..134b2001K} and books
\cite{2000oepn.book.....K, 2019ibfe.book.....B}.

  In this review we will focus on hydrogen(H)-deficient
CSPNe and their planetary nebulae and discuss our current understanding of
their formation and evolutionary connections. The paper is organized as follows 
In section \ref{sec:H-def-CSPNe-PNe} we describe the observed properties and classifications of both H-deficient CSPNe, and the subset of PNe that show  H-deficient ejecta. Then, in Section \ref{Born_Again} we discuss the so-called born again scenario for the formation of H-deficient CSPNe, and discuss its successes and shortcomings. Here we review different classifications and flavors of the born again scenario, and discuss their predicted chemical abundances.  Given their importance in the validation of the born again scenario, in Appendix \ref{bonafide} we provide a detailed discussion of the properties of confirmed bona fide born again stars. Having discussed the problems with the classical born again scenario we discuss, in Section \ref{progenitor_binary}, possible binary evolutionary channels for the formation of H-deficient CSPNe and discuss their own limitations. We close the review with some final comments in Section \ref{sec:conclusions}.
%\unskip

\section{Hydrogen deficient CSPNe and PNe}
\label{sec:H-def-CSPNe-PNe}

\subsection{Hydrogen deficient CSPNe}
\label{sec:sub_CSPNe}
\begin{figure}[H]
\includegraphics[width=10.5 cm]{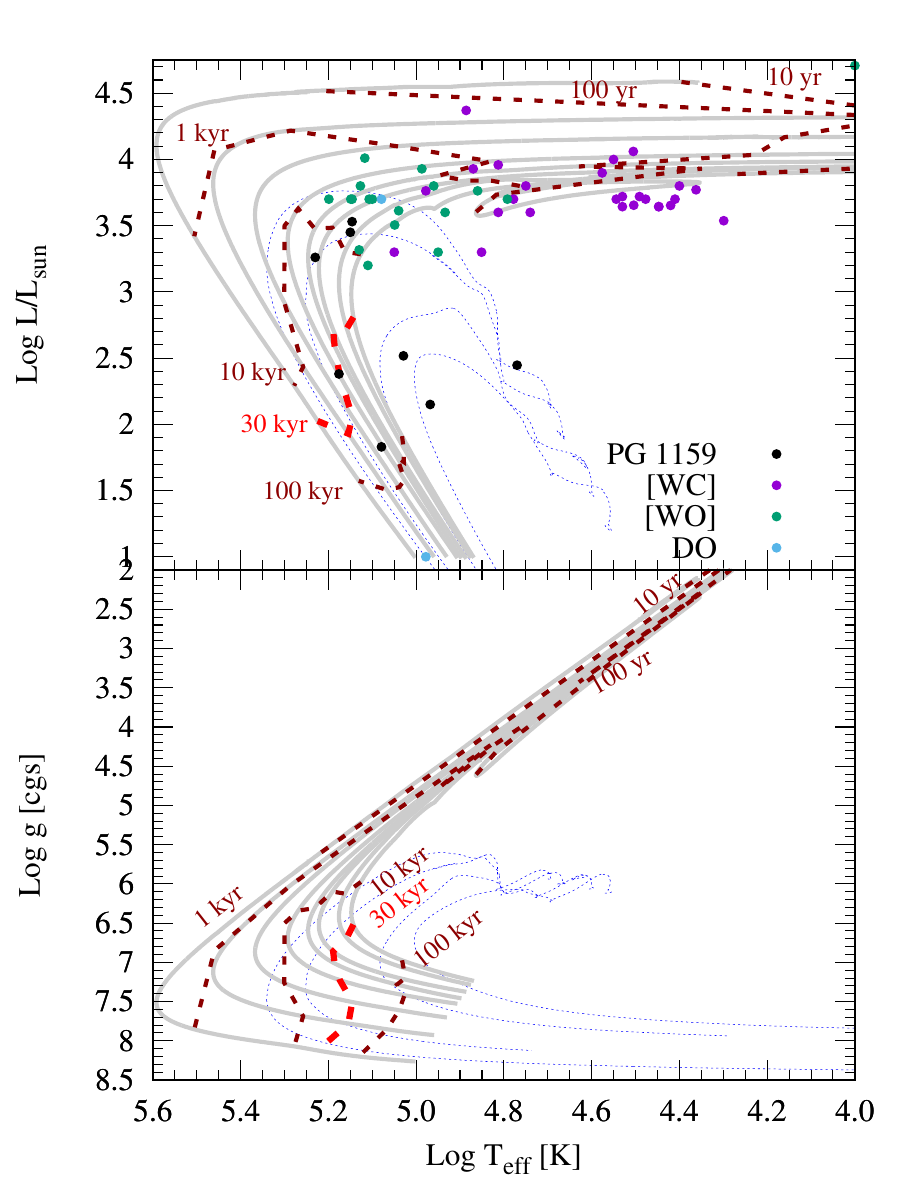}
\caption{ Upper panel: Location of the different H-deficient CSPNe in the HR diagram from the recent catalog of Weidmann et al. \cite{2020A&A...640A..10W} together with their spectral types. Note the dearth of [WR] stars at around $T_{\rm eff}\sim 50000$K. Grey lines indicate the location and evolution of  models of H-deficient CSPNe with their corresponding isochrones (dashed lines) \cite{2006A&A...449..313M,2006AA...454..845M,2007MNRAS.380..763M}.   Grey tracks correspond to post-VLTP sequences with remnant masses of (from right to left): 0.53, 0.542, 0.5647, 0.5886, 0.609, 0.6641, 0.7411 and 0.8697 $M_\odot$.  See Section \ref{sec:LateTP} for a discussion on the born again scenario and its different flavors.     Dotted blue tracks correspond to low-mass post merger models (He-core WD + low-mass CO-core WD) with masses 0.48, 0.55, 0.7 and 0.8  $M_\odot$ (from right to left). Lower panel: Evolution of the same models shown in the upper panel buy in a Kiel diagram.}
\label{fig:HR-Kiel}
\end{figure}

While the majority of CSPNe maintain their original H-rich
surface composition, the study of CSPNe exhibiting emission-line
spectra, which are very similar to those of massive Wolf-Rayet stars
with strong helium (He) and carbon (C) emission lines (i.e., spectral
types [WC] and [WO]\footnote{ The use of square brackets was proposed by  \cite{1981SSRv...28..227V}  to distinguish the Wolf-Rayet CSPNe  from their massive counterparts.  [WC] stars are also divided into
  "late'' (cool, spectral classes[WC 6]-[WC 12]) and ``early'' (hot,
  spectral classes [WC 4]-[WC 5]) spectral types and denoted by [WCL]
  and [WCE] respectively. All [WO] stars have "early'' spectral types
  (spectral classes [WO 1]-[WO 4]). In this paper we use [WCE] to
  include both [WC 4]-[WC 5] and [WO1]- [WO 4] spectral classes.}),
suggested the existence of H-deficient CSPNe stars
\cite{1954ApJ...119..243A, 1968nim..book..483A,
  1975ApJ...196..195H}. Later, the Palomar-Green survey revealed a new
spectral class of H-deficient stars, some of which are CSPNe, the
PG1159 stars, which are dominated by absorption lines of highly
ionized He, C, and O \cite{1985ApJS...58..379W}.  It is currently
understood that a sizable minority between one fifth and one third of
all CSPNe show a strong H-deficient surface composition
\cite{2020A&A...640A..10W}. The majority of these H-deficient CSPNe
exhibit spectra with strong C and He lines. Quantitative spectral
analysis of these stars determined that their surface is composed
predominantly by these elements
\cite{1989LNP...328..194W,1991A&A...247..476W,1994A&A...283..567L,1996A&A...312..167L,1998A&A...330..265L}. In
addition to He and C, both PG1159, [WC] and [WO] stars show an
important presence of oxygen(O). A couple of objects have been
discovered that display spectral characteristics intermediate to that
of PG1159 and [WC] spectral types, these stars are sometimes labeled
as [WC]-PG1159 \citep{2006PASP..118..183W}. Two well studied cases of
[WC]-PG1159 are the CS of Abell 30 and Abell 78, both which have been
previously linked to the so-called born again scenario
(\cite{1983ApJ...264..605I,1984ApJ...277..333I}, see Section
\ref{Born_Again}). The qualitatively similar He, C and O abundances
suggest that all these spectral types might be evolutionary connected.
 Fig. \ref{fig:HR-Kiel} show the typical surface luminosity and and effective temperatures of  different types of H-deficient CSPNe taken from  \cite{2020A&A...640A..10W} together with the evolutionary tracks of born again models (see Section \ref{Born_Again}).

In addition to the C-rich H-deficient group mentioned above, a
minority  of H-deficient CSPNe show He-dominated atmospheres, with traces of nitrogen (N).  The spectra  of these stars is mostly defined by
an almost pure He II absorption-line spectrum in the optical
wavelength range \cite{1986ASSL..128..323M,1991IAUS..145..375M, 1998A&A...338..651R}. The
spectral types of these He-dominated CSPNe are O(He), DO WDs, and at
least one [WN] (IC 4663, \cite{2012MNRAS.423..934M}). It is worth noting
that not all O(He) are CSPNe, and most DO WDs are not. Due to their
He-dominated atmospheres, and the presence of a measurable amount of N,
similar to that of the majority of R Coronae Borealis (RCrB) stars and extreme helium stars (EHe) stars \cite{1994Jeffery, 2008ASPC..391....3J}, the 
He-dominated CSPNe  have been linked to RCrB, EHe, and to the
occurrence of CO-core WD + He-core WD mergers \cite{2006ASPC..348..194R,2014AA...566A.116R}.
%se propone el vinculo con los merger de COWD
%y HeWD (pero ojo con la composicion del aPNe!!! no da ni por
%asomo. Merger en l marco de una envoltur comun parec la unic opcion)
%ver Hall & Jeffery 2016 con el tema del H en merger

Besides the previously mentioned groups there is an even smaller group
of H-deficient CSPNe that still display significant amounts of
H (larger than 10\% by mass fraction) and with different
spectral types. These are the CSs of Abell 48
([WN],\cite{2011MNRAS.414.2812D,2013MNRAS.430.2302T,2014MNRAS.440.1345F}),
PB 8 ([WN/C], \cite{2003A&A...403..659A,2010A&A...515A..83T}, and the
CSs of Sh 2-68, Abell 43, and NGC 7094, all classified as
``Hybrid PG1159'' stars, from now on H-PG1159 stars \cite{2006PASP..118..183W,2019MNRAS.489.1054L}. Good descriptions of the typical abundances in [WR] and PG1159 stars are discussed by \cite{2006PASP..118..183W,2014A&A...564A..53W,2015wrs..conf..253T}.

%%%PONER UNA DESCRPCION DETALLADA DE LO QUE PREDICEN LOS MODELOS DE
%%%BORN AGAIN, PARA LAS ABUNDANCIAS, DISCUTRI O, DISCUTIR NON , DISCUTIR N
%%% MENCONAR LA ABUNDANCIA DE H, CITAR TRABAJO PROCEEDING THIN NVELOPS

\unskip

\subsection{Pulsating H-deficient CSPNe (GW Vir)}\label{pulsations}
Some H-deficient CSPNe with PG1159 and [WCE] spectral types belong to
a small subgroup of variable stars known as GW Vir, after the
prototype of the class
PG\,1159-035 \cite{1979wdvd.coll..377M,1984ApJ...277..211G}. GW
Vir that are surrounded by a PN are also named variable planetary
nebula nuclei or PNNVs.  GW Vir exhibit multi-periodic
lightcurves that are attributed to non-radial pulsations in high
radial order g modes of low harmonic degree.  The typical pulsation
periods observed are in the range of 300s to 3000s, and are explained by
the action of the classical $\kappa-\gamma$ mechanism due to partial
ionization of C and O, which are extremely abundant in the envelope of
these stars \cite[see][ for a detailed discussion]{2019A&ARv..27....7C, 2020FrASS...7...47C}. The
existence of multiperiodic lightcurves in these stars opened the
possibility to use asteroseismological techniques to study the
properties of CSPNe. More specifically asteroseismology offers
unparalleled accuracy in the determination of
masses \cite{2007A&A...461.1095C,2007A&A...475..619C,2008A&A...478..175A,2008A&A...478..869C,2009A&A...499..257C,2021A&A...645A.117C,2024A&A...686A.140C},
and might even offer the opportunity to determine the internal rotation
profiles of CSPNe\cite{2009Natur.461..501C,2011MNRAS.418.2519C}. The
large number of periods found in GW Vir (usually about 20 frequencies
but up to 200 frequencies in the case of PG 1159-035) allows masses
to be determined to a precision of a few percent, exceeding what can
be determined by spectroscopic means in this complicated regime
\cite{2021A&ARv..29....4S}.  Moreover, pulsational properties of GW Vir
  have the potential to place constraints on the internal structure of PG1159 stars, and, consequently testing the validity of potential formation scenarios. This approach was explored by \cite{2005A&A...439L..31C} who tested the possibility of using mode-trapping features in the period spacings of GW Vir to put constraints on the properties of convective boundary mixing (CBM) during the core He burning stage.  More concretely \cite{2008ApJ...677L..35A} found that the mixture of positive and negative rates of change of the periods of the normal modes measured in the prototype star PG~1159-035 can be interpreted as a strong hint that PG1159 stars could be characterized by substantially thinner helium-rich envelopes than traditionally implied by the standard theory of the
formation and evolution of PG 1159 stars.

\subsection{Hydrogen deficient PNe}
\label{sec:H-def-PNe}
Almost all PNe have chemical compositions dominated by H with traces
of other chemical elements\cite{2022PASP..134b2001K}.  There exists,
however, a small group for which regions of H-deficient (He-dominated)
material have been discovered \cite{2002Ap&SS.279..171Z}. These PNe
are commonly termed "born again PNe'' for the suspected connection
with the born again stars\footnote{ Note that the term born again PNe is misleading, as during a born again event (section \ref{Born_Again}) the evolution of the central star might be fast enough so that the original PNe never ``dies''. This is actually what has been observed in Abell 58 (see Appendix \ref{bonafide}) were the CS returned to very high effective temperatures well before the old PNe dissapeared. Originally the term born again was coined with the idea of the rebirth of the central star \cite{1984ApJ...277..333I}.}.
To the best of our knowledge the group of PNe with known H-deficient
inner regions includes: Abell 30 \cite{1980Natur.285..463H}, Abell 78
\cite{1983ApJ...266..298J}, Abell 58 \cite{1987Msngr..50...14S}, IRAS
18333-2357 \cite{1989ApJ...338..862G,1989ApJ...346..803C},IRAS
15154-5258 \cite{1989A&A...218..267M}, the PNe around V4334 Sgr
(a.k.a. Sakurai’s Object) \cite{2002ApJ...581L..39K}, WR 72
\cite{2020MNRAS.492.3316G}, and HuBi 1 \cite{2022MNRAS.512.4003M}. All
these H-deficient PNe for which the CS has been spectroscopically
studied harbor in their center a H-deficient CS, which indicates
beyond reasonable doubt that a single event is responsible for
both deficiencies in H. Besides the common nature of their CSs these
born again PNe also share many other characteristics.  Most
importantly in almost all these PNe the outer H-rich PN is old,
circular (or near-spherical), and expanding at normal speeds while the
inner H-deficient regions expand much faster, have  axisymmetrical
shapes (disks and/or bipolar outflows), and display clear cometary knots
\cite{2002Ap&SS.279..171Z,2017MNRAS.470..626M,1996ASPC...96..193H,2020ApJ...903L...4R,2023ApJ...955..151R,2022MNRAS.512.4003M,2014ApJ...785..146H,2022ApJ...939..103R,2022MNRAS.514.4794R,2023A&A...677L...8T,2022ApJ...925L...4T,2020ApJ...904...34H}\footnote{An  exception to this case might be old PNe around HuBi 1 for which  Guerrero et al. \cite{2018NatAs...2..784G} reported a barrel-like
  structure with faint polar protrusions.}.
     Another interesting feature in H-deficient PNe comes from the so-called abundance discrepancy factors (ADFs\footnote{ADFs consists in the discrepancy found in ionized nebulae when deriving abundances from from recombination lines and collisionally excited lines\cite{1942ApJ....95..356W}. Typical values of the ADFs are around  2 to 3 in PNe \cite{2022PASP..134b2001K}.}). Although, to the best of our knowledge, the only H-deficient PNe with  measured ADFs are Abell 30 and Abell 58, values are very high in both cases. ADFs of 90 have been found for Abell 58 \cite{2008MNRAS.383.1639W} and, even higher values were derived for some H-deficient knots in Abell 30 (22, for J4\cite{2022RNAAS...6....4S}, 600 for J3, and 770 for J1\cite{2003MNRAS.340..253W})\footnote{ A compilation  PNe with measured ADFs is available on Roger Wesson’s website
  \url{https://nebulousresearch.org/adfs/}}. This feature is interesting both for the study of H-deficient PNe and of ADFs. In particular, a strong connection have been found, in standard PNe, between very high ADFs and the binarity of their central stars\cite{2018MNRAS.480.4589W}. Althouth it should be noted that the presence of H-deficient incluions has also been proposed as an explanation  for high ADFs\cite{2003IAUS..209..339L}. Regarding the possible binary nature of the CS of the 8 known H-deficient PNe described in this section, only for Abell 30 the possibility of a close binary CS has been suggested through the study of light curve brightness variations \cite{2020MNRAS.498L.114J}.

\section{The Born Again Scenario}
\label{Born_Again}
\begin{figure}[H]
\includegraphics[width=10.5 cm]{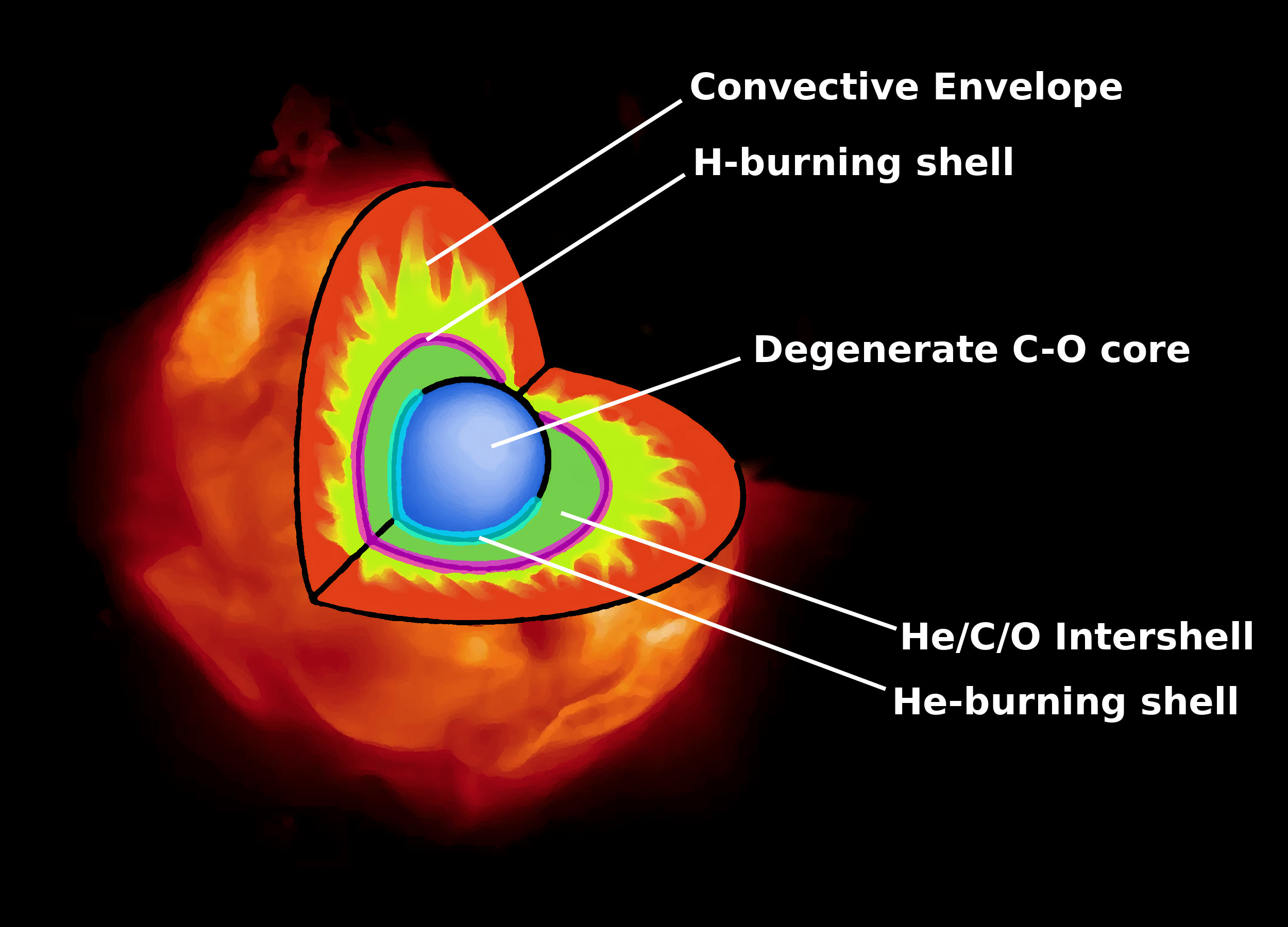}
\caption{Schematic diagram of the internal structure of an AGB star. Note that for the sake of clarity the diagram is severely not to scale. While outer convective envelope has a radius of the order of the Sun-Earth distance, the degenerate CO-core has a size comparable to that of Earth, and the outer He/C/O intershell  has a size similar to that of the icy planets of the Solar System. }
\label{fig:AGB}
\end{figure}   

\begin{figure}[H]
\includegraphics[width=11.5 cm]{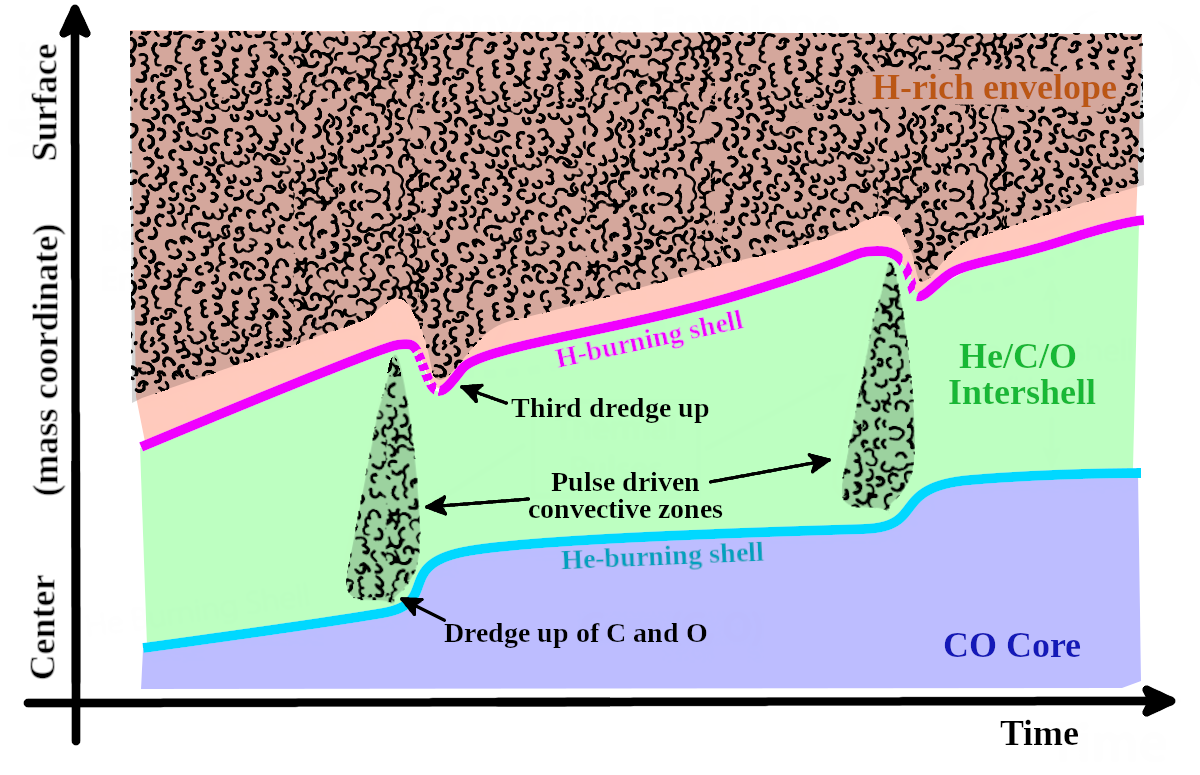}
\caption{Schematic Kippenhahn diagram of the internal structure of an AGB star during the thermal pulses. Colors are similar to those of Fig.\ref{fig:AGB}. }
\label{fig:Kipp-AGB}
\end{figure}

\subsection{The thermal pulses on the AGB}
\label{sec:agb}
One of the most accepted scenarios for the formation of H-deficient
CSPNe and WDs is the so-called born again scenario
\citep{1983ApJ...264..605I,1984ApJ...277..333I}. In order to
understand the essence of the born again scenario,  here we provide a brief
description of the immediate progenitors of those stars. We
refer the reader to \cite{2013sse..book.....K} for a detailed
explanation of the evolution of single low- and intermediate-mass
stars. Fig. \ref{fig:AGB} shows the internal structure of both low-
and intermediate-mass stars when they reach the AGB. After burning H in
the center during the main sequence, and then burning He in the
center, these stars have a well-layered structure.  At the center of this structure lies a CO core supported by
the pressure of degenerate electrons,  which is surrounded by a He-burning
shell. Above the He-burning shell there is an intershell region rich
in He, C and O, on top of which sits a H-burning shell, where H is
transformed into He by the CNO cycle. Finally, the H burning shell is
surrounded by an envelope composed mostly of unprocessed H and He, which
is what gives the AGB star its giant size.

The importance of the
structure of AGB stars for the study of H-deficient CSPNe lies in the
fact that, at the end of the AGB, the He-burning shell experiences a
series of instabilities where the energy released is temporarily
increased by about 5 orders of magnitude, or even more
\citep{2012sse..book.....K}. These instabilities are called thermal
pulses (or He-shell flashes) and have, as one of their most direct
consequences, the development of convective instabilities in the
stellar interior  (the so-called pulse driven convective zone, PDCZ). Fig. \ref{fig:Kipp-AGB} displays a schematic Kippenhahn diagram (i.e. mass coordinate $m(r)$ vs. time) of this phase. Cloudy regions in Fig. \ref{fig:Kipp-AGB} indicate convective regions where material is being mixed. Two main dredge up episodes might take place during this stage. One happens during the thermal pulse itself where the lower boundary of the PDCZ dabbles into the CO core of the star, dredging both C and O to the intershell region. The extent of this mixing depends both on the mass of the CO core and on the treatment of convective boundary mixing (CBM, a.k.a. ``overshooting''). More intense CBM leads to larger amounts of C and O being dredged up. In particular the more intense the CBM the larger the O enrichment of the intershell. The other dredge up episode, the so-called third dredge up \citep{2014PASA...31...30K}, might happen immediately after the thermal pulse, when the envelope expands reacting to the thermal pulse. There the outer convective zone can penetrate into the He-, C- and O-rich intershell, polluting the H-rich envelope with the material from the intershell. Whether third dredge up occurs or not, again, depends both on the mass of the core and the treatment of CBM, as well as on the metallicity of the star  \citep{2014PASA...31...30K}. Low-mass AGB stars (e.g. initially $\sim 1 M_\odot$ stars, with cores of $\sim 0.53 M_\odot$) will not experience third dredge up under standard CBM assumptions, while more massive stars will experience progressively stronger third dredge up episodes\citep{2014PASA...31...30K}. Besides enriching the envelope in He, C and O, third dredge up plays a critical role in the production of heavy elements through slow neutron capture process \citep{2014PASA...31...30K}. Conditions at the bottom of the convective envelope during third dredge up episodes allow for the partial mixing of $^{12}C$ and H, which can later be burned to produce $^{13}C$ during the interpulse phase. When the next thermal pulse happens  $^{13}C$ will be engulfed by the PDCZ and burnt with He, leading to the creation of free neutrons that can feed the  slow neutron capture process.

 At the same time the star is undergoing thermal pulses it is subject to very strong winds ($\dot{M}\sim 10^{-8}\mbox{--}10^{-4}\, M_\odot$/yr).
The accepted view of AGB winds is that they are driven by radiation pressure on solid-state particles (dust
grains) formed in the outer regions of the AGB atmospheres that are levitated by pulsation. Depending on the composition
of the atmosphere these dust particles are either made of silicates (when C/O$< 1$ by number) or amorphous carbon (when C/O$> 1$ by number).
All AGB stars are initially of spectral type M (with C/O $\simeq 0.5$ by number), but when third dredge up is active, newly produced carbon may eventually lead to a ratio of C/O $> 1$ in the atmosphere (carbon star, spectral type C)\footnote{ In the transition from spectral type M to C, S-stars are formed
  ($0.5< $  C/O $ < 1$) }. This transition from M-stars into carbon stars can be critical for the AGB winds, as silicate dust is made from elements that are not produced in low-mass stars
and consequently winds in this stage will depend on the original metallicity of the star. Conversely, carbon dust is based on primary carbon produced by the AGB star and C-dust driven winds will be mostly independent on the original metallicity of the star. A detailed review of AGB winds and their modeling can be found in  \cite{2018A&ARv..26....1H}. In addition to this,  \cite{2002A&A...387..507M} showed that, as soon as the the star becomes a carbon star and carbon  rich molecules
are formed in the atmosphere, the star becomes cooler and more extended, increasing the intensity of the AGB winds.
The main consequence of these strong winds is that the H-rich envelope is finally reduced below the critical mass that allows for a giant configuration. When this happens the star will depart from the AGB and start evolving, first as a post-AGB star, and latter as a hot pre-WD. If conditions are appropriate a PN can be formed during this transition (see panel a in Fig. \ref{fig:HRDiagram})

\subsection{Thermal pulses after the AGB (late thermal pulses)}
\label{sec:LateTP}

\begin{figure}[H]
\includegraphics[width=\columnwidth]{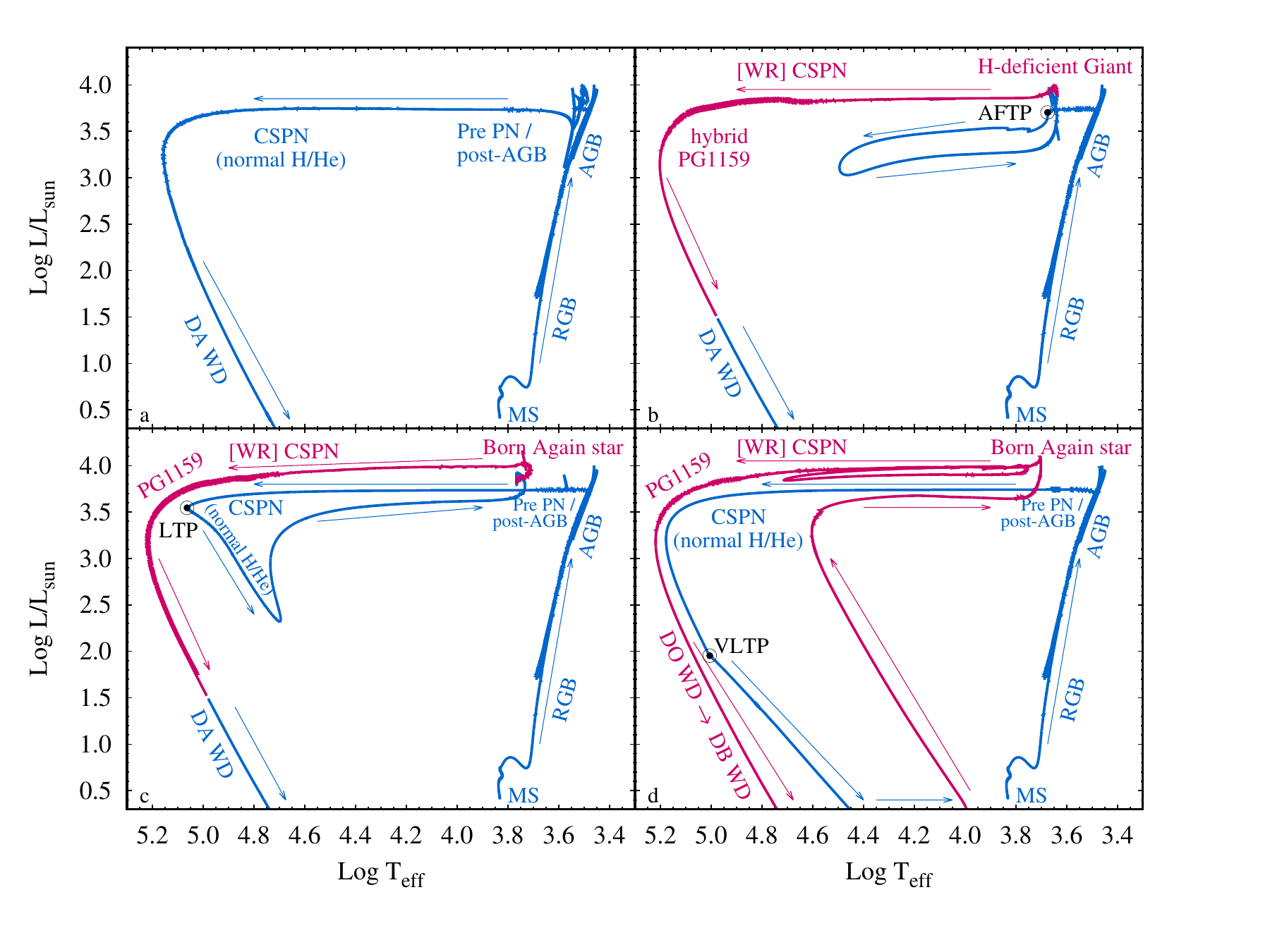}
\caption{ Evolution on the HR diagram of an initially $1.25 M_\odot$  star from the main sequence (MS) to the final WD phases. Particular emphasis is done on the departure from the AGB into the post-AGB/pre-PN phases, and the formation of a PN and its central star. Arrows indicate the direction of the evolution. Blue line indicates the regions in the HR diagram where the model shows a H-rich surface composition, while magenta lines indicate a strong H-deficient surface composition.  Panel a: canonical post-AGB  evolution. Panel b: formation of a H-deficient CSPN after an AGB final thermal pulse (AFTP). Panel c: formation of a H-deficient CSPN after a late thermal pulse (LTP). Panel d: formation of a H-deficient CSPN after a very late thermal pulse (VLTP). }
\label{fig:HRDiagram}
\end{figure}

\begin{figure}[H]
\includegraphics[width=12.5 cm]{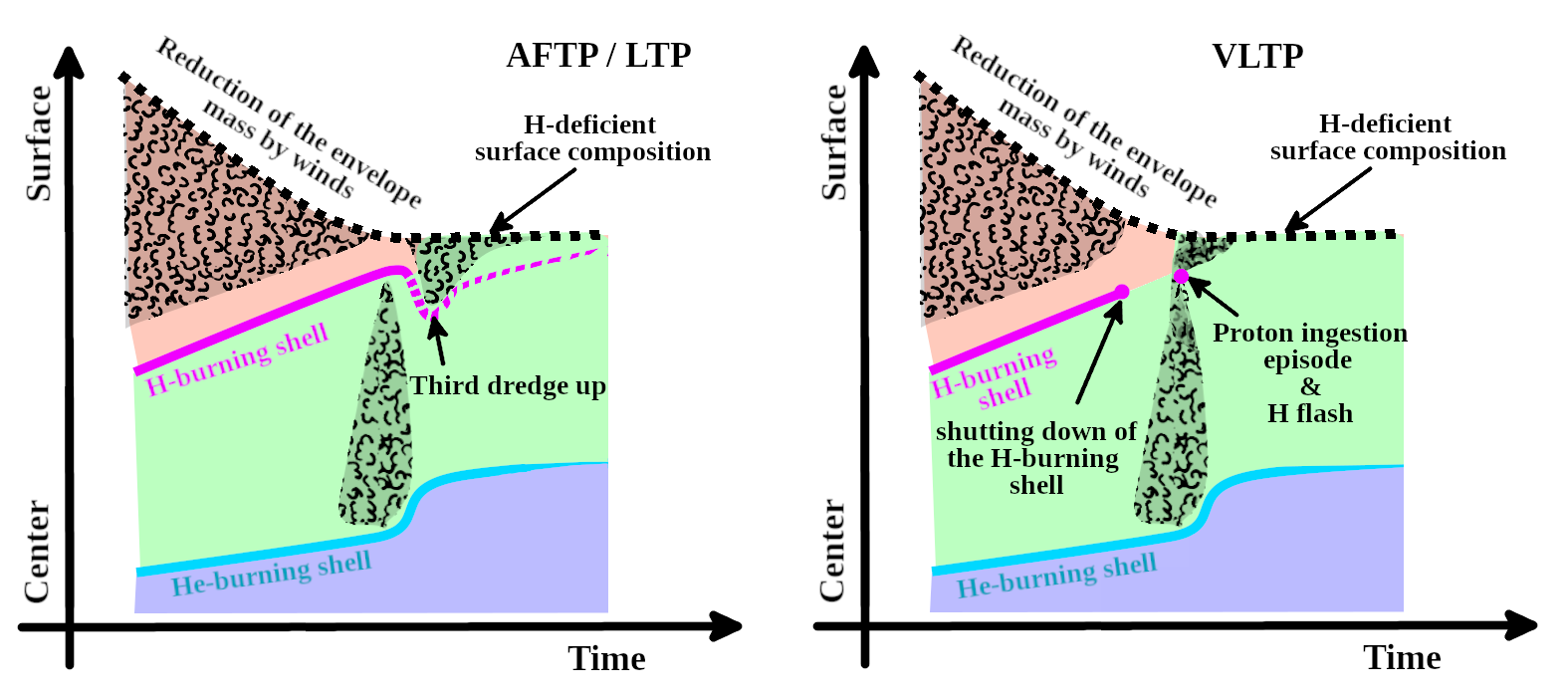}
\caption{Schematic Kippenhahn diagrams of the internal structure of a post-AGB star that undergoes an AFTP/LTP (left) or a VLTP (right).}
\label{fig:Kipp-pAGB}
\end{figure}   

Of particular
interest to us is the fact that the thermal pulses described in  Section \ref{sec:agb}  can occur when
most of the envelope of the star has been removed by the stellar
winds, and the star is evolving towards the white dwarf phase
\citep{1979A&A....79..108S}  (see Fig. \ref{fig:HRDiagram}). These thermal pulses are then called 
late thermal pulses. In these cases, the sudden injection of energy by
the thermal pulse pushes the star back into a giant configuration,
giving rise to what have been called born again AGB stars
\citep{1984ApJ...277..333I} (see panels c and d in Fig. \ref{fig:HRDiagram}).
 For this reasons late themal pulse scenarios are collectively mentioned as the born again scenario.
Depending on the exact moment in the
post-AGB evolution in which the late thermal pulse occurs, different
outcomes are possible. And several of them can lead to the formation
of a star with a H-deficient surface
composition. Ref. \cite{2001Ap&SS.275....1B} offers a useful
classification of late thermal pulses into three categories: 1) AGB
final thermal pulse  (AFTP, panel b in Fig. \ref{fig:HRDiagram}) which occur just as the star is coming out
of the AGB, and the H/He-envelope mass is about of $M_{\rm env}\sim
10^{-2}\hbox{--}10^{-3} M_\odot$\footnote{The envelope masses
  associated to each type of late thermal pulse given here are for
  $M_{\rm CSPN}\simeq 0.6 M_\odot$ central stars. More massive
  remnants will have less massive H/He envelopes at each phase. For a
  $M_{\rm CSPN}\gtrsim 0.8 M_\odot$ AFTPs will happen with envelopes
  of $M_{\rm env}\sim 10^{-5}\hbox{--}10^{-4} M_\odot$, LTPs will have
  $M_{\rm env}\sim 10^{-4}\hbox{--}10^{-5} M_\odot$ and VLTPs $M_{\rm
    env}\lesssim 10^{-5} M_\odot$}; 2) late thermal pulse  (LTP, panel c in Fig. \ref{fig:HRDiagram}) when
the thermal pulse occurs during rapid evolution at constant
luminosity, and the mass of the H/He envelope is $M_{\rm env}\sim
10^{-4} M_\odot$; and 3) very late thermal pulses  (VLTP, panel d in Fig. \ref{fig:HRDiagram} ) when the
thermal pulse occurs once the H-burning shell layer has decreased its
intensity by more than an order of magnitude ($M_{\rm env}\lesssim
10^{-4} M_\odot$), the luminosity of the star has fallen by more than 2
or 3 orders of magnitude, and the star is entering the WD phase.  Each flavor of the thermal pulse leads to different chemical abundances in the photosphere of the CSPNe. Examples of the predictions of different models available in the literature are shown in Table \ref{tab:models}.

If the dredging-up processes that bring material to the
surface of the AGB star are active (the so-called third dredge
  up, 3DUP), the AFTP and LTP scenarios predict the formation of  H
deficient stars once the star returns to the giant branch and the
outer convective zone deepens.  As shown in Fig. \ref{fig:Kipp-pAGB} (left panel) if third dredge up is active, after the thermal pulse (either LTP or AFTP) convective motions in the envelope will penetrate into the He-, C- and O- rich intershell. However, contrary to what happens on the AGB, now the H-rich envelope has a very low mass, as most of it was removed by winds on the AGB. The mass of the H-rich envelope is now comparable (AFTP) or smaller (LTP) than the amount of mass dredged up from the intershell. This leads to a strong He, C and O enrichment of the envelope of the star that can be directly observed at the photosphere.
Due to the difference in the mass of
the remaining H-rich envelope, AFTP and LTP scenarios predict different
final abundances for the star. While the LTP predicts final abundances
of H of $X_H\lesssim 0.05$ (in mass fraction), AFTP predicts
significantly higher abundances of $X_H\gtrsim 0.1$. In both cases the
predicted final surface abundances are extremely enriched in He, C and
O. It is worth emphasizing that the AFTP and LTP scenarios predict these
abundances only if 3DUP is active. If this is not the case surface
abundances remain unchanged and the result will be an CSPNe with an
H/He ratio close to solar.

 On the contrary the VLTP scenario (see panel d in Fig. \ref{fig:HRDiagram}) always predicts an extremely
H-deficient surface composition. This is because, in a VLTP, H is
violently burnt \citep{1995LNP...443...48I}, instead of being just diluted. In a VLTP the thermal pulse happens once the H-burning shell has already died out and the star is already entering the WD phase (see right panel in Fig. \ref{fig:Kipp-pAGB}).  Due to this the entropy barrier that normally prevents the contact between the PDCZ and the H-rich envelope is strongly reduced \citep{1976ApJ...208..165I}, and the PDCZ is able to penetrate the H-rich envelope, dragging inwards the H-rich material into the hot He-burning regions. H is then violently  burnt with $^{12}$C, and later also with  $^{13}$C and $^{14}$N, leading to a H-flash and a strong depletion of the H content of the star. A detailed description of these processes can be found in \cite{2006A&A...449..313M}. In addition to the formation of a strongly H-deficient surface composition, the VLTP might also lead to  the formation of heavy elements through intermediate neutron capture process \cite{2011ApJ...727...89H}.
Since the early works \cite{1979A&A....79..108S,1983ApJ...264..605I,1984ApJ...277..333I}
several authors have computed the formation of H-deficient CSs after a
VLTP. In order to compute the simultaneous burning and mixing of the
H-rich material into the He- and C-rich intershell it became a
standard practice in 1D stellar evolution models to adopt a diffusive
convective scheme
\cite{1995LNP...443...48I,1999A&A...349L...5H,2003ApJ...583..913L,2006MNRAS.371..263L,2005A&A...435..631A,2006A&A...449..313M,2018NatAs...2..784G,2020A&A...638A..30B}. The
predicted abundances vary depending on the details of the evolutionary
models but some generalities can be described. First and foremost it
should be noted that the main composition of the resulting H-deficient
star mostly reflects the previous He- and C- dominated composition of
the intershell \cite{1995LNP...443...48I}. This composition varies
with the number of thermal pulses and the initial mass of the star
\cite{2016A&A...588A..25M}. It is important to
point out that the O abundance is strongly dependent on the
assumptions about CBM 
at the bottom of the thermal pulse driven convective zone
\cite{1999A&A...349L...5H}. In order to reproduce the O-enriched
observed in PG1159 and [WR] stars ($\sim 10$\% to $\sim 15$\% by mass
fraction) mixing beyond the bottom of the pulse driven convective zone
needs to be added and calibrated \cite{1999A&A...349L...5H,
  2016A&A...588A..25M}.  Table \ref{tab:models} shows surface
abundances of different post-VLTP models once they become H-deficient.
  Fig. \ref{fig:HR-Kiel} shows post-VLTP evolutionary tracks and isochrones from \cite{2006AA...454..845M,2007MNRAS.380..763M}. When compared with H-rich central stars of similar mass  (e.g. figure 8 in \cite{2016A&A...588A..25M}) massive  H-deficient CSPNe evolve slower, and stay brighter for a longer time, than their H-rich counterparts.

A different classification of late-helium flashes was proposed by
\cite{2006MNRAS.371..263L}. In their classification, Type I models
leave the AGB during the hydrogen-burning phase and continue like that
until they reach the WD cooling stage and never become
H-deficient. Type II models corresponds exactly to the VLTP
classification described above, and always become H-deficient. Types
III and IV models experience a helium shell ﬂash after leaving the
AGB, but do it while the H-burning shell is active, so that H is not
burnt. The difference between types III and IV models is based on the
effective temperature of the star when the helium shell ﬂash occurs
($T_{\rm eff} > 30 000 K$, Type III, and $T_{\rm eff} < 30 000 K$,
Type IV). Consequently, all Type III for models correspond to LTP
cases, while Type IV corresponds to LTP and AFTP cases.
Finally\footnote{Lawlor \cite{2006MNRAS.371..263L} also defines four more type for those high
  metallicity sequences that, in their sequences, evolve away from the
  AGB before reaching the phase of thermal pulses.}, Type V models leave the
AGB during a helium shell ﬂash and type VI models leave the AGB during
quiescent helium burning.  Some type V and VI models become slightly
H-deficient. In addition to these types, \cite{2023MNRAS.519.5373L} added a transition type II/III  for those models that reach peak helium-burning luminosity when $T_{\rm
  eff > 100 000 K}$ and the H-burning shell is decreasing its
luminosity but its still quite active. In his work \cite{2023MNRAS.519.5373L} the author reports
that in their computations (in which no CBM is applied to the PDCZ) type II/III
models consume about half the H.

\startlandscape
\begin{table}[H]
  \centering
\caption{Typical surface abundances (by mass fractions) of post-VLTP, post-LTP and
  post-AFTP evolutionary models that lead to the formation of
  H-deficient central stars (during the PG1159 and [WR] stages). $M_i$ and $M_f$ indicate the initial (zero age main sequence) and final (WD stage) masses of each evolutionary sequence.
  Note that models with $Z_0=0.02$ were computed with slightly different assumptions for CBM than
  $Z_0=0.01$ and $Z_0=0.001$ models. \label{tab:models}}
%  \begin{tabularx}{\textwidth}{rc|cccccccc|ccc}
  \begin{tabularx}{19cm}{rc|cccccccc|ccc}

  \toprule
 \textbf{$M_f$} & flavor & \textbf{H} & \textbf{$^4$He} & \textbf{$^{12}$C} & \textbf{$^{13}$C} & \textbf{$^{14}$N} & \textbf{$^{16}$O} & \textbf{$^{20}$Ne} & \textbf{$^{22}$Ne}& \textbf{$M_i$} &  \textbf{$Z_0$} & \textbf{Ref.}\\
 \textbf{[$M_\odot$]} &   &             &       &       &       &       &       &   &    & \textbf{[$M_\odot$]} &  &  \\  \midrule
0.530 & VLTP & $9 \times 10^{-3}$   & 0.33 & 0.39 & 0.051             & 0.0192            & 0.17 & $1.6 \times 10^{-3}$  & 0.0194 & 1.0  & 0.02 & \cite{2006AA...454..845M} \\
0.542 & VLTP  & $6.9 \times 10^{-3}$ & 0.28 & 0.41 & 0.051             & 0.0178            &  0.213 & $1.6 \times 10^{-3}$  & 0.0192 & 1.0  & 0.02 & \cite{2006AA...454..845M} \\
0.561 & VLTP  & $1 \times 10^{-4}$   & 0.33 & 0.32 & 0.053             &  0.0289           &  0.232 &$1.6 \times 10^{-3}$  & 0.0275 & 1.8  & 0.02 & \cite{2007MNRAS.380..763M} \\
0.564 & VLTP  & $1.1 \times 10^{-3}$  & 0.39 & 0.27 & 0.048            &  0.0265           &  0.217 &$1.7 \times 10^{-3}$  & 0.0380 & 2.2  & 0.02 & \cite{2006AA...454..845M} \\
0.584 & VLTP  & $\sim$0              & 0.40 & 0.31 & 0.055              &  0.0362         & 0.170 &  $1.6 \times 10^{-3}$  & 0.0295 & 2.5  & 0.02 & \cite{2006AA...454..845M} \\
0.609 & VLTP  & $\sim$0              & 0.50 & 0.35 & $3.1 \times 10^{-3}$ &$2.1\times10^{-3}$ & 0.103 & $1.6 \times 10^{-3}$ & 0.0355 & 3.05 & 0.02 & \cite{2006AA...454..845M} \\
0.664 & VLTP  & $\sim$0              & 0.47 & 0.33 & 0.019               & 0.0128           & 0.126 & $1.9 \times 10^{-3}$  & 0.0328 & 3.5  & 0.02 & \cite{2006AA...454..845M} \\
0.741 & VLTP  & $\sim$0              & 0.48 & 0.34 & $7.1 \times 10^{-3}$ & $2.5\times10^{-5}$ & 0.139 & 0.0026 & 0.027 & 3.75  & 0.02 & \cite{2007MNRAS.380..763M}\\
0.870 & VLTP  & $\sim$0              & 0.54 & 0.30 &  $9.9 \times 10^{-3}$ &  $8.3 \times 10^{-3}$ & 0.0938 & 0.0061 & 0.0171 & 5.5  & 0.02  & \cite{2006AA...454..845M} \\ \midrule
  0.480  & VLTP  & 0.030             & 0.52 & 0.30 & 0.041                & 0.016           & 0.086 & 9$\times$10$^{-5}$ & 1$\times$10$^{-3}$ & 0.8 & 0.001 & \cite{2018NatAs...2..784G} \\
  0.537  & VLTP  & 5$\times$10$^{-3}$ & 0.48 & 0.29 & 0.069                & 0.065          & 0.091 & 9$\times$10$^{-5}$ & 9$\times$10$^{-4}$ & 0.9 & 0.001 & \cite{2018NatAs...2..784G}  \\
  0.535  & VLTP  & 5$\times$10$^{-5}$ & 0.40 & 0.34 & 0.074                & 0.056          & 0.116 & 9$\times$10$^{-4}$ & 0.01 & 1.0 & 0.01 & \cite{2018NatAs...2..784G}  \\
  0.552  & VLTP  & 5$\times$10$^{-3}$ & 0.38 & 0.37 & 0.059               & 0.028           & 0.145 & 9$\times$10$^{-4}$ & 0.01 & 1.1 & 0.01 & \cite{2018NatAs...2..784G}   \\
  0.567  & VLTP  & 5$\times$10$^{-6}$ & 0.41 & 0.33 & 0.067               & 0.047            & 0.133 & 9$\times$10$^{-4}$ & 0.01 & 1.25 & 0.01 & \cite{2018NatAs...2..784G},Fig. \ref{fig:HRDiagram}  \\ \midrule  
0.565  & LTP  & 0.049             & 0.33 & 0.36 &  $4.8 \times 10^{-7}$   &   $7.2 \times 10^{-5}$ & 0.21 & 1.7$\times$10$^{-3}$ & 0.038 & 2.2 & 0.02 & \cite{M3B-PhD} \\
0.589  & LTP  & 0.040             & 0.31 & 0.40 &  0.0138                &   $1.4 \times 10^{-3}$ & 0.20 & 1.7$\times$10$^{-3}$ & 0.020 & 2.7 & 0.02 & \cite{2005BAAA...48..185M, M3B-PhD} \\
0.567  & LTP  & $9 \times 10^{-3}$ & 0.42 & 0.43 &   $5.5 \times 10^{-7}$  &   $1.5 \times 10^{-4}$ & 0.12 & 9$\times$10$^{-4}$ & 0.01 & 1.25 & 0.01 &  Fig. \ref{fig:HRDiagram}\\

\midrule
 0.566  & AFTP  & 0.137             & 0.39 &\multicolumn{2}{c}{0.36}  &   $6 \times 10^{-4}$ & 0.104 & \multicolumn{2}{c}{8.5$\times$10$^{-3}$}  & 1.25 & 0.01 & \cite{2019MNRAS.489.1054L},Fig. \ref{fig:HRDiagram} \\
 0.550  & AFTP  & 0.197             & 0.45 &\multicolumn{2}{c}{0.30}  &   $1 \times 10^{-4}$ & 0.056 & \multicolumn{2}{c}{7.8$\times$10$^{-4}$} & 1.0 & 0.001 & \cite{2019MNRAS.489.1054L} \\
 0.594  & AFTP  & 0.219             & 0.45 &\multicolumn{2}{c}{0.26}   &  $7 \times 10^{-4}$ & 0.055 & \multicolumn{2}{c}{0.021}  & 1.5 & 0.001 & \cite{2019MNRAS.489.1054L} \\ 
  \bottomrule 
  \end{tabularx}
\end{table}
\finishlandscape

Lawlor \citep{2021MNRAS.504..667L,2023MNRAS.519.5373L} presented a
detailed grid of stellar evolution models of LTP events.  Their models
cover a wide metallicity range ($Z=0.0015\hbox{--}0.03$) but are only focused in
LTP events in low-mass progenitors $(0.9 M_\odot<M_i<2M_\odot)$. As
these simulations do not include CBM most
models never become H-deficient.  Only a few of their TP models (their
"type II/III" models), those that erupt at the highest temperatures
(very close to the boundary with VLTP cases ), and with the smallest
radii become modestly enriched in He, C, N, and O. This may be
relevant for CSPNe that are only modestly H deficient.  On the
contrary, models that include CBM, and favor
the development of 3DUP
\cite{2003IAUS..209..101B,2005A&A...440L...1A,2005BAAA...48..185M}
lead to much stronger H deficiencies (H $\sim 0.05$ by mass fraction).
H deficiency is attained after the return to the AGB, when third
dredge up develops in the post-LTP giant
\cite{2003IAUS..209..101B}. In these cases He, C and O abundances are
very similar to those predicted by the VLTP case, as the models is
displaying its previous intershell abundances at the photosphere (see Figs.\ref{fig:Kipp-AGB} and \ref{fig:Kipp-pAGB}). The
main difference is in the isotopic $^{13}C/^{12}C$ ratio, as VLTPs
lead to a strong production of $^{13}C$ and in many cases also
$^{14}N$. Table \ref{tab:models} shows surface abundances for some
post-LTP models that became H-deficient.

The number of AFTP models in the literature resulting in H-deficient
surface compositions is very limited
\cite{2001Ap&SS.275...15H,2006MNRAS.371..263L, 2019MNRAS.489.1054L}.
For those models that undergo third dredge
up\cite{2001Ap&SS.275...15H, 2019MNRAS.489.1054L} the final surface
composition of H is between $X_{\rm H}\sim 0.1$ and $X_{\rm H}\sim
X^\odot_{\rm H}$ depending on how massive the remaining H-rich
envelope is at the moment of the helium flash. As in LTP cases, the
final O abundances are strongly dependent of the assumed CBM during
the previous AGB evolution (see Table \ref{tab:models}).

Besides 1D stellar evolution simulations, in the last two decades
3D hydrodynamical simulations of the H entrainment in the He-shell
flash convective zone during a VLTP event became possible
\citep{2011ApJ...727...89H,2014ApJ...792L...3H,2015ApJ...798...49W}. These
simulations show a very rich complexity of hydrodynamic processes that
cannot be captured by 1D stellar simulations of the event.  Unfortunately,
3D hydrodynamical simulations are very demanding. Current simulations
of the convective reactive ingestion of protons into the He-burning
shell during a VLTP can only be performed for a very short physical
time, a snapshot in the development of the VLTP. Simulations by Herwig
et al. \citep{2011ApJ...727...89H,2014ApJ...792L...3H} correspond to
physical times of about a day in the life of the star. Consequently,
the development of the violent H-flash cannot be followed and
simulations need to be started from an artificial background model
where the H-flash develops as soon as possible. In spite of this,
3D-hydrodynamical simulations are very informative, inspired by 3D
hydrodynamical simulations it was suggested \cite{2011ApJ...727...89H} that the
splitting of the convective zones created by the violent H-burning
during H-entrainment could be delayed, leading to observable signatures
in the nucleosynthesis by intermediate neutron capture
processes\footnote{Intermediate neutron capture processes happen at
  neutron densities intermediate to those of the classical slow ($N_n
  \sim 10^7$--$10^{11}$ cm$^{-3}$) and rapid neutron capture
  processes ($N_n \sim 10^{20}$
  cm$^{-3}$)\cite{1957RvMP...29..547B}.}.  More interestingly, the
simulations of the H ingestion flash
performed by \cite{2011ApJ...727...89H,2014ApJ...792L...3H}  found the formation of global non-spherical oscillations that
are sustained by individual ignition events of H-rich fluid
pockets. This results is particularly interesting in view of the
plethora of non-spherically symmetric features found around
H-deficient CSPNe \cite{2014ApJ...785..146H} (see Section
\ref{sec:H-def-PNe}).  Moreover, 3D hydro simulations have been used
to inform simplified 1D models of convective mixing.  For example, \cite{2021MNRAS.504..744S} 3D-hydro simulations were used to
construct a 1D advective two-stream model for the computation of
detailed nucleosynthesis.

 The final evolutionary state of stars undergoing an AFTP, LTP or VLTP deserves some comment. It is a well known fact that once stars enter the WD cooling track, winds fade and gravitational settling takes over, leading to the formation of the observed almost pure (H or He) atmospheres of WDs \cite[see][for a detailed review of WD evolution]{2010A&ARv..18..471A}. Even tiny amounts of diluted H ($M_H\simeq 10^{-14}M_\odot$)  can lead to the formation of a WD with  pure H atmosphere (DA WD) \cite{2020A&A...633A..20A}. In the AFTP and LTP cases where a H-deficient CSPNe was formed after 3DUP, the H diluted into the deeper parts of the envelope is later burned as the star contracts again to the WD cooling track leading to a decrease in the H content of the WD. Once winds stop and  gravitational settling takes over both post-LTP and post-AFTP stars are expected to form DA WDs. In the case of post-LTP stars DA WD with very thin H envelopes are expected, with  H-contents as low as $M_H \simeq 10^{-6}\hbox{--}10^{-7} M_\odot$ \cite{2005A&A...435..631A,2005BAAA...48..185M} (see Fig. \ref{fig:HRDiagram}). The final fate of post-VLTP remnants is less understood. It has long been argued that post-VLTP stars will for H-deficient WDs (DO, DB, DC and DQ WDs, see Fig. \ref{fig:HRDiagram})\cite{2005A&A...440L...1A}. However the complete destruction of H seems to be very difficult in stellar models, either thorugh burning or winds on the [WR] stage.   For the case of post-VLTP stars with masses $M_f \simeq 0.6 M_\odot$, \cite{2017ASPC..509..435M} obtained DA WDs with H contents as low as $10^{-11} M_\odot$. Better models are needed to clarify whether post-VLTP stars end up as DA or non-DA WDs.

\subsection{Advantages and successes of the born again scenario}
Although the return stage to the AGB after the thermal pulse is
extremely fast in evolutionary terms, some stars have been identified
going through these very fast stages. Four CSPNe have been
measured to change their surface properties ($T_{\rm eff}$, $L_\star$,
$g$ and even composition) in the short timescale of a few years or
decades corresponding to late thermal pulses (FG~Sge, V605~Aql,
V4334~Sgr, and Hen 3-1357, see Appendix \ref{bonafide}). Of these
stars, V605~Aql and V4334~Sgr have been identified as VLTP cases,
while FG~Sge and Hen 3-1357 seem to be LTP cases.  These born again
stars not only offer a confirmation that late thermal pulses actually
occur in nature, but also demonstrate the link between AGB stars and
CSPNe. Moreover, they allow testing different particularities of
stellar evolution models and their nucleosynthesis
\citep{2005ARA&A..43..435H} (see Appendix \ref{bonafide}). 
V4334 Sgr, FG Sge, and also V605 Aql demonstrate that the born again
scenario actually leads to the formation of H-deficient CSPNe, and
also, that it can lead to the formation of [WO] CSPNe
\cite{1997AJ....114.2679C,2002Ap&SS.279..183D,2004A&A...426..145L,2006ApJ...646L..69C}. Moreover,
the born again scenario offers a natural qualitative explanation for
the similar abundances found in [WC], [WO], and PG1159 spectral types,
and also to the luminosity and photospheric temperatures and gravities
of these stars, linking them to a natural evolutionary post-AGB
sequence, see lower panel of Fig.  \ref{fig:HRDiagram}.

May be the most appealing characteristic of the born again scenario
is that its occurrence is statistically unavoidable
\cite{1979A&A....79..108S,1983ApJ...264..605I,1984ApJ...277..333I}. Given
that the occurrence or not of the born again scenario (and its
flavors, VLTP, LTP, AFTP) is only a consequence of the relation
between the timing of thermal pulses and the timing of the removal of
the envelope of the AGB star \cite{2018A&ARv..26....1H}, and these two
clocks run independent from each other, for a given initial mass and
metallicity it is expected that a sizable fraction of all single
stars will undergo a born again event. This fraction is expected to be
somewhere around $\sim 20$\% \cite{1984ApJ...277..333I}. In the
absence of any theoretical argument suggesting that the departure from
the AGB always happens at the right phase to avoid a late helium
flash, one should come to terms with the idea that late thermal pulses
happen in a sizeable minority of single post-AGB stars, and form H-deficient
CSPNe through this channel. Current estimations of the multiplicity of
low- and intermediate-mass stars \cite{2023ASPC..534..275O} indicate
that the close binary fraction ($a<10$au) is of about $\sim 20\% $ for
stars with initial masses of $M_i\simeq 1M_\odot$ and rises to $\sim
55\%$ for stars with $M_i\simeq 6M_\odot$. This means that between
80\% and 45\% of all intermediate and low mass stars evolve mostly
unperturbed by possible companions. The large number stars evolving in
isolation or in wide binaries, together with the sizeable fraction of
stars for which departure from the AGB will result in a born again
event, is a compelling argument in favor of the idea that many
H-deficient CSPNe and post-AGB stars have a born again origin.

Besides its ubiquity, one of the most successful features of the born
again scenario is its ability to reproduce the rare He, C, and O
atmospheres of both [WC], [WO] and PG1159 stars
\cite{1999A&A...349L...5H,2001Ap&SS.275....1B}. Qualitatively, it is
very unlikely to have layers in the interior of star where He, C, and
O can coexist. The existence of He, C and O in the surface of [WC],
[WO] and PG1159 stars links them almost exclusively to the intershell
of AGB stars. A possible exception to this might be the outcome of the
merger of two low-mass CO-core WDs or the merger of a low-mass CO-core WD and a
more massive He-core WD \cite{2022MNRAS.511L..60M}. However, in these cases
the formation of a PNe with normal H/He compositon would be imposible to
explain. Conversely, the born again scenario naturally explains the
presence of a PNe with a normal H/He composition. Additionally the late
thermal pulse scenario could help understand how bipolar nebulae could
form around isolated stars
\citep{2014ApJ...792L...3H,2014ApJ...785..146H,2018Galax...6...79V}.
The late thermal pulse scenario is a key piece of the current
understanding about the formation of around of 20\% of the CSPNe,
which show H deficiency
\citep{2011A&A...526A...6W,2020A&A...640A..10W}.

 A final argument in favor of the late flasher scenarios (AFTP, LTP, VLTP) cn be made from ther predictions for the final WD stage. The existence of single WDs with extremely thin H envelopes (of the order of $M_H \simeq 10^{-14} M_\odot$) has been invoked to explain the the spectral evolution that WDs seem to experience as they evolve \cite{2010A&ARv..18..471A, 2020MNRAS.492.3540C,2024Ap&SS.369...43B}. More specifically the recent study by \cite{2020MNRAS.492.3540C} indicates that 60\% of WDs must have a H contents larger than $M_H = 10^{-10} M_\odot$, another 25\% have H contents in the range $M_H = 10^{-10} M_\odot\hbox{--}10^{-14} M_\odot$, and 15\% have  H contents lower than $M_H = 10^{-14} M_\odot$. The existence of DA WDs with very thin H envelopes has also been inferred from asteroseismological studies of variable DA WDs \cite{2012MNRAS.420.1462R, 2022MNRAS.511.1574R}. Such low-H content WDs cannot be explained by canonical single stellar evolution (panel a, Fig. \ref{fig:HRDiagram}) which predicts that  WDs should be formed with a total H content of $M_H \simeq 10^{-3}\hbox{--}10^{-5} M_\odot$, depending on the stellar mass and metallicity of the progenitor \cite{2015A&A...576A...9A}. As discussed in Section \ref{sec:LateTP} these single DA WDs with thin H contents are naturally explained by LTP and VLTP scenarios \cite{2017ASPC..509..435M}.

\subsection{Shortcomings of the born again Scenario}

Despite its many successes, the born again scenario faces serious
challenges as it is unable to explain many observed features in
H-deficient CSPNe and their PNe. In fact, none of the bona fide late
helium flahers discussed in Appendix \ref{bonafide} evolved exactly
like models predicted. For example, while the $T_{\rm eff}$ evolution
of Hen 3-1357 is well reproduced by an LTP in a low-mass remnant
($0.53<M_{\rm CSPN}/M_\odot<0.56$, \cite{2017MNRAS.464L..51R}) the
surface gravity of Hen 3-1357 is sistematically higher than that
predicted by LTP models
\cite{2017MNRAS.464L..51R,2021MNRAS.504..667L}. Similarly, while the
pre-outburst and post-outburst evolution of V605 Aql and V4334 Sgr can
be reproduced by standard VLTP models of different masses
\cite{2007MNRAS.380..763M,2011ApJ...743L..33M} no single model is
able to simultaneusly reproduce all these observables. Moreover, as
mentioned before, both the 1921 spectra of V605 Aql and the 1996
pectra of V4334 Sgr show much higher He abundances and very little C
in comparison with the intershell abundances of AGB stars and VLTP
models.

With no hope of being exhaustive in this section we will mention some
of the shortcomings and failures of born again models.  In this
connection we can mention that the properties of the H-deficient
material ejected around some H-deficient CSPNe are at odds with our
current understanding of the the born again scenario. The abundances
on the H-deficient knots around Abell 30 \cite{2003MNRAS.340..253W}
and Abell 58 \cite{2008MNRAS.383.1639W} were found to have $C/O<1$ and
the presence of substantial quantities of neon (Ne) ($\sim 34$\% and
$\sim 13$\% respectively\footnote{In the case of Abell 30 this is true
  only for one of the knots analyzed as the Ne abundance is not the
  same for all of them.}), respectively at variance with the
predictions of the born again scenario.  It should be noted, however, that according to \cite{2021MNRAS.503.1543T}  if the carbon trapped into dust grains is taken into account then the  C/O ratio of the H-deficient ejecta  is larger than 1, and in agreement with the born again scenario.
Another problem for the born again scenario are the He/C/O abundances of
the H-deficient knots of Abell 58 and Abell 30 which are not in agreement
with the current abundances of their CSs which show typical PG1159
He/C/O abundances (and show [WCE]-PG1159 spectral types). It should be
noted however that, if shocks are indeed the dominant excitation
mechanism, the current nebular abundance estimates
\cite{1996ApJ...472..711G,2008MNRAS.383.1639W} should be
revisited.  Moreover studies of Ne and O abundances in the H-deficient ejecta in Abell 30 \cite{2003MNRAS.340..253W}, Abell 58 \cite{2008MNRAS.383.1639W} and Abell 78\cite{2023MNRAS.526.4359M} found a very high Ne/O ratios not in agreement with the predictions of the born again scenarios. In
addition to the abundance inconsistency, there is a clear discrepancy
between the geometries of the outer and inner nebular regions of both
Abell 30 and Abell 78.  The outer H-rich shells are ellipsoidal and
expand at $\sim 40$ km s$^{-1}$, while the H-deﬁcient knots detected
have much larger velocities of up to 200 km s$^{-1}$. The central
parts in Abell 30 and Abell 78 were imaged by the Hubble Space
Telescope and revealed structures distributed on an equatorial plane
and polar features \cite{2014ApJ...797..100F,2022MNRAS.514.4794R}. Both the strong
axisymmetrical shape of the inner regions and their high Ne abundances
have been used to link these objects to posible binary evolution
channels (see Section \ref{binary_results}).

Another feature that hints that that the born again scenario cannot be
the only formation scenario for H-deficient CSs is the detection of
close binary stars with H-deficient compositions. Jacoby et al
\cite{2020MNRAS.498L.114J} report the presence of light curve
brightness variations having a period of $P_{\rm orb} = 1.060$ d in
Abell 30, hinting at the possibility of a close binary CS. A similar
discovery was reported by \cite{2015MNRAS.448.1789M} who found a
signiﬁcant periodic variability hinting at a close binary ($P_{\rm orb} = 4.04$ d)
near the [WO] CS of NGC 5189.  The presence of a
close binary companion within the born again scenario is not easy to
reconcile. For that situation to happen it would be necesary that the
VLTP happens immediately after the common envelope event, as otherwise
the H-burning shell would still be active and not VLTP would
occur. Yet, if a VLTP where to occur in a close binary system the
expansion following the VLTP would lead to the engulfing of the
companion and the destruction of the close binary.  Similarly, for the
same reasons it is very difficult to imagine a born again origin for
the only known PG1159 star in a close binary
\cite{2006A&A...448L..25N,2009JPhCS.172a2065S}. These observations
indicate that there might be a binary evolution channel able to
produce the typical [WR]/PG1159 surface compostions. We will see in
the next sections that this suggestion is not devoid of problems. It is
worth noting that the case of NGC 246 does not pose a problem for the
born again scenario as the members of the triple system are relatively
distant \cite{2014MNRAS.444.3459A}.

A very well known problem for the born again scenario arises from the
lack of intermediate spectral types in Wolf-Rayet cental stars ([WC
  5-7]). This can be appreciated in Fig \ref{fig:HR-Kiel} as a dearth
of stars at about $T_{\rm eff}\simeq 50000$K. Due to the
expected evolution on the HR diagram after the return to the AGB (see
Fig. \ref{fig:HRDiagram}), one would expect that the second
contraction of the, now H-deficient, CSPNe would follow the spectral
sequence [WCL]$\rightarrow$ [WCE]$\rightarrow$PG1159, and then due to
the action of gravitational settling into DA WD or DO WD stars\footnote{DA WDs show almost pure H atmospheres with, at most, traces
of other elements (DAZ spectral type) and represent the vast majority of WDs. 
WDs with He dominated atmospheres are a minority and sometimes known as
non-DA WDs. Non-DA WDs are divided into several subclasses: DO WDs (45000 K $<  T_{\rm eff} <$  200000 K), DB WDs (11000 K $<  T_{\rm eff} <$  30000 K), and DC, DQ, and DZ types ($T_{\rm eff} < 11000$ K) depending 
which trace elements are visible in their spectra. See \cite{2010A&ARv..18..471A} for a detailed review. }
depending on the amount of H remaining in the envelope
\cite{2005A&A...435..631A,2005A&A...440L...1A,2005BAAA...48..185M,2022ApJ...930....8B}).
Consequently,
under the born again scenario the absence of intermediate spectral
types in [WR]-CSPNe is difficult to explain. Whether this is feature
is real or a consequence of the different spectral lines adopted for
the determination of the surface properties of early and late [WR]
stars is not known.

Moreover it has been argued that the there is a discrepancy between
the abundances of C and He of early and late [WC] types. While [WCL]
where determined to have similar He and C mass fractions (He/C/O
0.4/0.5/0.1,
\cite{1993AcA....43..329L,1994A&A...283..567L,1996A&A...312..167L,1998A&A...330..265L}
similar to those observed in PG1159 stars, the earlier [WCE] were
determined to have twice as much He than C by mass fraction [WCE]
He/C/O 0.6/0.3/0.1
\cite{1997A&A...320...91K,2001Ap&SS.275...41K,2015wrs..conf..253T}. A
similar situation happened for transition types [WC]-PG1159 (Abell 30
\& Abell 78) He~0.65, C=0.20...0.30, N=0.015, O=0.15 . However this
discrepancy has been questioned by some authors, finding similar C and
He abundances for [WCE] and [WCL] (He/C/O ~ 0.45/0.45/0.1
\cite{2003IAUS..209..243C,2007ApJ...654.1068M}.  To date, [WCL] have
not been reanalyzed systematically with with improved line-blanketed
stellar atmosphere models. The latter determinations of abundances in
[WR] stars with modern atmosphere models seem to sugget that both
groups might have similar abundances (see Table \ref{tab:abu_Crich}).

\begin{table}[H] 
\caption{Recent determinations of surface abundances in [WCL], [WCE], and [WO] CSPNe.\label{tab:abu_Crich}}
%\newcolumntype{C}{>{\centering\arraybackslash}X}
\begin{tabularx}{\textwidth}{lccc}
\toprule
\textbf{Name}	& \textbf{Spectral type}& \textbf{[H/He/C/N/O/Ne]}	& \textbf{Reference}\\
 	&      & \textbf{(mass fractions)} & \\
\midrule
HuBi 1 & [WC10] &  0.01...0.05/0.33/0.5/0.01/0.1/0.04 & \cite{2018NatAs...2..784G}\\
SwSt 1 & [WC9/10] &  0 /0.42/0.50/0.05/0.03/0.02 &\cite{2020MNRAS.498.1205H}\\
NGC 40 & [WC8] &  0 /0.57/0.4/1.e-3/0.03/<0.03 & \cite{2019MNRAS.485.3360T}\\\hline
NGC 2371 & [WO1] & 0.71/0.20/0.001/0.06/0.03 & \cite{2020MNRAS.496..959G} \\ 
NGC 6905 & [WO2] &  <0.05/0.55/0.35/7e-5/0.08/0.02 & \cite{2022MNRAS.509..974G} \\
NGC 1501  & [WO4] & -/0.6/0.3/0.15/-/- & \cite{2022MNRAS.517.5166R} \\\hline 
Abell 30 & [WCE]-PG 1159 &  -/0.63 /0.20 / 0.15 / 0.015/- & \cite{2012ApJ...755..129G}\\
Abell 78 & [WCE]-PG 1159 &  -/0.55 /0.30 / 0.10 / 0.015 / 0.04 & \cite{2015ApJ...799...67T}\\
\bottomrule
\end{tabularx}
\end{table}

Another problem for the born again scenario comes from different
estimations of the spatial distribution of normal CSPNe, PG1159, and
[WR] stars. In their work, Weidmann et
al. \cite{2011A&A...526A...6W,2020A&A...640A..10W} found that the
galatic latitude distribution of H-rich and H-deficient CSPNe is
different, with H-deficient CSPNe closer to the galactic plane. That
would be an indication that the progenitors masses and ages of both
populations are different, with the progenitors of H-deficient stars
more massive and younger than their H-rich counterparts. While this is
not in strong contradiction with the born again scenario it would
require the frequency of late thermal pulses to increase with mass,
while the opposite is hinted by models, where the shortening of
post-AGB timescales with mass \cite{2016A&A...588A..25M} happens much
faster than the shortening of the interpulse cycle with mass, making
late flashes less probable. Moreover, a more serious problem has been
suggested by a new catalog of distances to PNe based on Gaia paralaxes
\citep{2024arXiv240304606H}. By deriving distances in a homogeneous
way for 2211 objects the authors find that [WR] CSPNe are closer to
the Galactic plane than H-rich CSPNe. Most importantly, the authors
find that [WR] CSPNe are closer to the galactic plane than other
H-deficient CSPNe (such as PG 1159, O(He) and DO WDs), with the latter
having similar distribution to H-rich CSPNe. If this is confirmed it
would suggest that [WR] CSPNe have more massive progenitors than
PG1159 CSPNe (see also \citep{2013RMxAA..49...87P}, making the
evolutionary connection impossible for most of these stars. This later
result is at variance with the earlier work of
\cite{2000A&A...362.1008G} who analysed the observational data for PNe
with H-deﬁcient CSPNe and concluded that it suggested the evolutionary
sequence the general evolutionary sequence
[WCL]$\rightarrow$[WCE]$\rightarrow$PG 1159. Noteworthy, this work
also concluded that the observed parameters of PNe were not consistent
with the theoretical models of the born again scenario available at
that time \cite{2000A&A...362.1008G}. A reanalysis  in
light of updated models of the born again scenario is necessary.

Finally, it is clear that the born again scenario is unable to explain
the properties of the He-rich sequence of CSPNe (O(He), DO, and some
[WN], see Table \ref{tab:abu_Herich}). Note that while it is expected
that DO stars are formed from PG1159 stars, the timescales for
gravitational settling turning a PG1159 star into a pure DO WD are
much longer than any expected lifetime for a PNe
\cite{2005A&A...435..631A}. Consequently, those DO that are CSPNe
cannot descend from PG1159 CSs.
While objects such as PB 8 and Hen 2-108 have H/He abundances not far
from those of H-PG1159 stars and might be  evolutionarily connected (see
Table \ref{tab:abu_Herich}, although C abundances greatly differ) the
situation is very different for the "pure'' [WN] stars (IC 4663 and
Abell 48). Both IC 4663 and Abell 48 show extreme He abundances,
similar to those of the O(He) and DO CSPNe, but very different from those
predicted by all flavors of the born again scenario. These abundances
are very similar to those observed in RCrB stars \cite{1996PASP..108..225C}, and
consequently have been connected to possible merger episodes  (see Section \ref{progenitor_binary}).

\begin{table}[H] 
  \caption{Recent abundance determinations  [WN], O(He), DO, and Hybrid-PG1159 (H-PG1159) CSPNe.\label{tab:abu_Herich}}
%  \label{tab:abu_Herich}
%\newcolumntype{C}{>{\centering\arraybackslash}X}
\begin{tabularx}{\textwidth}{lccc}
\toprule
\textbf{Name}	& \textbf{Spectral type}& \textbf{[H/He/C/N/O/Ne]}	& \textbf{Reference}\\
	& & \textbf{(mass fractions)} & \\
\midrule
IC 4663 & [WN 3] & <0.02/0.95/<0.001/0.008/0.0005/0.002 &\cite{2012MNRAS.423..934M} \\
Abell 48 & [WN 5] &  0.1/0.85/0.03/0.05/<0.006/- &\cite{2013MNRAS.430.2302T}\\
PB 8 & [WN 6/WC 7] &  0.4/0.55/0.01/0.01/0.01/- &  \cite{2019MNRAS.485.3360T} \\
Hen 1-108 & [Of/WN 8] & 0.50/0.48/$7\times 10^{-5}$/$1.4\times 10^{-5}$/$5.7\times 10^{-5}$ & \cite{2023MNRAS.522.1049M} \\\hline
 K 1-27 & O(He) & 0.046/0.933 /$5.6\times 10^{-4}$/0.0132/$5\times 10^{-5}$/0.005 & \cite{2014AA...566A.116R} \\
 LoTr 4  & O(He) & 0.12/0.871/<$7.2\times 10^{-4}$/0.0079/<$1.9\times 10^{-4}$/<0.001 & \cite{2014AA...566A.116R} \\\hline 
Abell 43 & H-PG1159 & 0.25/0.46/0.27/0.0026/0.0044/0.012 & \cite{2019MNRAS.489.1054L}\\
NGC 7094 & H-PG1159 & 0.15/0.52 / 0.31 / $3\times 10^{-4}$ / 0.0033 / 0.0019 & \cite{2019MNRAS.489.1054L}\\\hline
KPD 0005+5106 & DO & <0.025/0.977/0.01/0.0025/0.004/0.01 & \cite{2015AA...583A.131W} \\
\bottomrule
\end{tabularx}
\end{table}

\section{PNe progenitors and binary stellar evolution}\label{progenitor_binary}

It is a well known fact that binarity in low- and intermediate-mass
stars is not as ubiquitous as in massive stars
\citep{2010ApJS..190....1R,2023ASPC..534..275O}. However, the
multiplicity fraction changes dramatically in the range of masses
expected for the progenitor stars of PNe.  For example, for
KGF main sequence stars (initial masses $0.75 M_\odot<M_i
<1.5M_\odot$) the fraction of stars in multiple systems is reported to
be between 42\% and 50\% \cite{2023ASPC..534..275O}, but it rises to about 70\% for A stars
($\sim 2M_\odot$), and then to to $\sim 80$\% for late B stars ($3
M_\odot<M_i <5M_\odot$), and to $\sim 90$\% for early B stars ($5
M_\odot<M_i < 8M_\odot$). Nevertheless, while wide binaries might
affect the shape of the PNe they are unable to alter the evolution and
lead to true binary evolution channels. For example, FGK main sequence
stars follow a broad lognormal separation distribution peaking near $a
= 40$ au \citep{2010ApJS..190....1R,2014AJ....147...86T}, so most low-mass
stars will follow single star evolution paths, regardless of the
existence of a companion\footnote{The evolution might involve,
  however, the interaction with substellar companions.}. This said,
the close binary fraction among low- and intermediate-mass stars is
not negligible. In fact the close binary fraction ($a<10$ au) is
between 20\% and 25\% 
for FGK main sequence stars, rises to $\sim 37$\% for A-type stars,
and then to $\sim 46$\% and $\sim 55$\% for late- and early-type B
stars respectively \citep{2010ApJS..190....1R,2014AJ....147...86T,2023ASPC..534..275O}.

In the following paragraph we will very briefly describe some of the
intricacies of binary stellar evolution, the reader interested in the
many physical processes and details of binary stellar evolution is
referred to to specialized books and reviews on the topic
\citep[e.g.][]{2006epbm.book.....E,2014LRR....17....3P,2017PASA...34....1D,
  2023pbse.book.....T}.  One of the main mechanisms of interaction in
a binary system is the transfer of mass (and angular momentum) between
the components of the system. The most important modes of mass
transfer are: Bondi–Hoyle–Lyttleton (BHL) wind accretion
\citep{1944MNRAS.104..273B}, the wind accretion through Roche lobe
overflow (WRLOF) \citep{2007ASPC..372..397M}, stable mass transfer by Roche
Lobe Overflow (RLOF) \citep{1971ARA&A...9..183P}, and unstable mass
transfer with the consequent formation of a common envelope (CE)
\citep{1976IAUS...73...75P}. In addition to these processes
interaction through tides and magnetic fields can also play a role in
the orbital evolution of the system \cite{1975A&A....41..329Z,2022ApJ...933...25P,2024arXiv240710573E}. All these phenomena have as a
consequence the transfer and/or loss of angular momentum between the
components and change the orbital parameters of the system.  The two
dominant parameters in defining the evolution of the system are the
distance ($a$) and the mass ratio $q=M_2/M_1$ between the
components. The mass transfer can occur on dynamic, thermal, or
nuclear time scales, depending on the mass ratio and the response of
the radius of the donor star to mass loss. The type of mass transfer
that occurs in a particular system depends primarily on whether the
stellar envelope of the donor is convective or not (which determines
how the star radius reacts to mass loss), and how the ratio of masses
between both stars is (which determines how the size of the Roche
lobes changes as mass is transferred/lost from the system).

When the mass transfer occurs on a very short a time scale the
envelope of the accreting star is pushed away from the so-called
thermal equilibrium (steady state). Because of this it expands and
might form a common envelope around both stars.  A common envelope will lead to
rapid decay of the orbit, due to drag forces and the transfer of
energy and angular momentum to the common envelope
\citep{1976IAUS...73...75P}. If the envelope is ejected the result it
will be a close binary surrounded by the ejected material. If, in
addition, the temperature evolution of the remnant, and the gas
dispersion times, are favorable this will lead to the formation of a
planetary nebula with a close binary inside.

The critical mass ratio for mass transfer to be unstable is not well
determined. Recent studies indicate that, for conservative mass
transfer, mass transfer would only be unstable for values of
$q>1.5\hbox{--}2$ \citep{2015MNRAS.449.4415P}, greater than the
threshold value of $q\sim 0.8$ that emerged from more idealized
studies \citep{2000eaa..bookE1624E}. When the mass transfer is stable
the system widen or shrink depending on whether the system loses some
of its mass and angular momentum during mass transfer
\citep{2014LRR....17....3P}. In relation to the formation of planetary
nebulae,  \cite{2018MNRAS.473..747C} studied the orbital
evolution of binary systems where the donor is an AGB star. There,
using simulations of stars of 1$M_\odot$ with $q=2$ and $q=10$
(i.e. companions of 0.5 and 0.1 $M_\odot$), and initial separations
between 3 and 10 AU (periods from 5 to 26 years), found that systems
with a separation greater than $\sim 6$ AU (periods of $\sim 12$ years
or $\sim 4380$ days) end in more distant orbits as a consequence of
BHL mass transfer while those systems with shorter distances end up in
tighter orbits as a consequence of WRLOF mass transfer. So, systems at
initial distances smaller than 6 AU would end up approaching each other as a
consequence of the WRLOF, and could later evolve through RLOF and CE
stages depending on the initial distance.

An additional comment regarding the impact of binaries on the
formation of PNe refers to the time scales corresponding to contraction to
the WD  stage. Ref. \cite{2013MNRAS.435.2048H} showed that
models of stars undergoing rapid mass loss (with envelopes far from the thermal
equilibrium) evolve much faster than the classical
models where the stellar envelope is lost by stationary winds on time
scales of thousands of years.

\subsection{Proposed evolutionary channels for H-deficient CSPNe}\label{binary_results}
The appeal of binary channels for the formation of
H-deficient CSs is based on two very different circumstances. On the
one hand binarity opens a large parameter space for the initial
configuration of the system that might help to explain the diversity
of CSPNe and PNe properties.  This is in line with the studies of  Abell 30 \cite{2020MNRAS.498L.114J} and NGC 5189 \cite{2015MNRAS.448.1789M} which suggest that both PN might  harbor a close binary CSs.  Moreover, the high ADFs observed in both Abell 30 and Abell 58 \cite{2008MNRAS.383.1639W, 2022RNAAS...6....4S, 2003MNRAS.340..253W}, together with the clear connection between strong ADFs and binarity\cite{2018MNRAS.480.4589W} make binarity and appealing possibility. 
On the other hand, the lack of a
quantitative understanding of unstable mass transfer and common
envelope events
\cite{2013A&ARv..21...59I,2014LRR....17....3P,2017PASA...34....1D,2020rfma.book..123J}
allows for broader speculation than single star evolution models. In
this section we will only discuss some scenarios proposed in the
literature for the formation of CSPNe with no hope of being
exhaustive. The reader interested in the relevance and variety of
binary formation channels of both CSPNe and PN is referred to
specialized reviews and books
\cite{2019ibfe.book.....B,2017NatAs...1E.117J,2020Galax...8...28J,2020rfma.book..123J}.

One of the most appealing binary scenarios for the formation of
H-deficient CSs is the merger of (low-mass) He-core WD with a
more massive CO-core WD. It has been long shown that CO-core WD + He-core WD merger can
explain the formation of the vast majority of RCrB stars
\cite{1984ApJ...277..355W,1996PASP..108..225C,2002MNRAS.333..121S,2007ApJ...662.1220C}. RCrB
stars have typical abundances of $X_{\rm H}<0.05$, $X_{\rm He}\sim
0.98$ and traces of C, N, O below 1\% by mass fraction
\cite{2000A&A...353..287A,2002AJ....123.3387D}. This abundances are in
qualitative agreement with what is expected from CO-core + He-core WD mergers,
where the less massive He-core WD (composed of almost pure He and traces of
N) is tidally disrupted and poured on top of the CO-core WD. The ignition
of He in a series of shell flashes in the He-mantle is able to produce
significant quantities of C and O while the star expands to its giant
(RCrB stage)\cite{2002MNRAS.333..121S}. Due to the obvious abundance
similarities between the abundances of RCrB stars and those of O(He)
and some [WN] CSPNe (IC 4663 and Abell 48, see Table \ref{tab:abu_Herich}) it seems natural
to link them evolutionary.  Moreover stellar structure models of RCrB
stars predict that these objects will finally contract at constant
luminosity crossing the CSPNe region of the HR diagram, before finally
becoming WDs
\cite{1987A&A...185..165W,2002MNRAS.333..121S,2019ApJ...885...27S}.
Thus abundances of the CSPNe and post-merger evolutionary tracks,
suggest the connection Merger $\longrightarrow$
RCrB $\longrightarrow$ EHe $\longrightarrow$ [WN] $\longrightarrow$
O(He) $\longrightarrow$ DO. The main shortcoming of this scenario is
that the PNe surrounding all these He-dominated stars show normal H
dominated compositions
\citep{1994A&A...286..543R,1996A&A...310..613R,2009A&A...496..139G,2012MNRAS.423..934M,2014MNRAS.440.1345F,2022MNRAS.511.1022D,2023FrASS..1022980M}. A
relatively natural alternative would be to suggest that the merger
takes place inside a common envelope which, once ejected, forms the PNe.
Under this scenario one would expect the PN to be much more massive
than normal PNe and strongly axisymmetrical, which does not seem to be
the case \citep{1994A&A...286..543R,1996A&A...310..613R}.

A similar reasoning can be applied to the low-mass CO-core WD + He-core WD
merger (or double low-mass CO-core WD merger) channel proposed in
\cite{2022MNRAS.511L..60M}. While it can explain the abundances
observed both in PG1159 and [WR] CSs the existence of a
normal (H-rich) PN around the merger product could only be explained
if the merger happens during a common envelope event. As in the
previous case one would expect the PN to be much more massive than
normal PNe and strongly axisymmetrical.

Other binary scenarios have been proposed in the literature
specifically tailored to explain certain properties of H-deficient
CSPNe \cite{2008ASPC..391..209D}. In this connection we can mention
the engulfment/merger scenarios aimed at explaining the so-called
double-dust chemistry of PNe\footnote{The presence in the PNe of
  C-rich dust located inside a shell of O-rich dust (silicates).} and
the H-deficiency of the CSPNe \cite{2002PASP..114..602D}.  The authors
proposed that, as the star reaches the AGB it engulfs a companion,
which spirals in within the AGB envelope, leading to an increase in
the AGB mass-loss rate. Then, the entire envelope is expelled as
O-rich gas, and C-enhancement in the inner layers is induced by shear
mixing caused by the engulfment of the companion. It is then assumed
that the enhancement of mass loss also leads to the formation of a
H-deficient central star. This scenario seems unlikely for several
reasons. Besides the obvious speculative nature that prevents
quantitative comparison, it has some problems even at the qualitative
level. Most importantly, as most of the envelope would be ejected at a
single event, we should expect most of the mass to be located close to
the star, and a very massive PNe around the star. As mentioned before,
those old nebula around H-deficient ejecta do not agree with these
characteristics.  It should be noted, however, that some fraction of the mass of the outer nebula could be in neutral form and hence missed from emission line studies. In addition to the low mass of the PNe, PNe around H-deficient stars are not
particularly axisymmetrical. Moreover, the innermost region of the
envelope, where the H-burning layer is located, and also the He-buffer
below it are very tightly bound to the core of the AGB star and should
be very difficult to be ejected during the merger process. In addition
mixing might also be very difficult due to the presence of a strong
entropy barrier between the envelope and the core
\cite{1976ApJ...208..165I}.  Ejection of any remaining H-envelope
through winds is also highly unlikely \cite{2017ASPC..509..435M}.

Another of the scenarios mentioned in the literature concerns the
so-called ``two-common envelope scenario''
\cite{2008ASPC..391..209D}. In this scenario it is assumed that a
final thermal pulse might happen to one of the components of an
already existing close binary system (after a previous, first, common
envelope event). In such a scenario, the final thermal pulse would
trigger a second common envelope event which should lead to the merger
of the two components. It is speculated then, that the companion
material would provide the H-rich gas that, together with C-rich wind
of the newly formed [WCL] star, would lead to the formation of PAHs.
Again, besides the speculative nature of the scenario, there are some
key points that might even be problematic on a qualitative
level. First, and foremost, it is not easy to envisage a situation in
which a VLTP might happen after a first common envelope. For a final
thermal pulse to happen the He-rich intershell needs to grow to a
critical value, this happens due to the CNO fusion at the H-burning
shell.  In single stars, VLTPs happen when the AGB envelope is reduced
, in thermal equilibrium, beyond a critical value through the action
of winds\cite{2016A&A...588A..25M}. As the star contracts towards the
WD cooling track H-burning shell continues burning H and increasing
the mass of the He-rich intershell\cite{2012sse..book.....K}.
%{\bf  Hacer un diagrama de estas posibilidades?}
This would not be the case after the first common envelope event that
forms the initial close binary.  The fast removal of the envelope
would lead the envelope away from thermal
equilibrium\cite{2013MNRAS.435.2048H} allowing (albeit not forcing)
the removal of the envelope well beyond the aforementioned critical
value. After the fast ejection of the envelope the remaining envelope
is either above or below the critical envelope value. If it is well
below, then the remaining H-rich envelope will be very likely too thin
to allow the He-rich intershell to grow to a critical value and
produce a VLTP, unless the situation is extremely finely tuned. If it
is well above, then the star will still have a giant configuration and
the companion has either been engulfed into the envelope of the
primary, and merged, or it is still in a relatively wide orbit
($a\gtrsim 1$AU). In both cases the star will continue evolving as a
single star. To the best of our understanding the only viable
possibility for this scenario to happen is that the first common
envelope event leads to the ejection of most of the envelope of the
AGB star but not all, close to the critical envelope mass needed for
the star to have a giant configuration, that the orbit shrinks but not
into a close binary, so that it can still contain the contracting
primary in thermal equilibrium. This might allow the development of a
VLTP and a second common envelope event\cite{2018MNRAS.475.4728B}. It
remains to be seen if such fine tuning of the ejected envelope is
statistically plausible. In this case, as the material remaining in
the star undergoing the VLTP is not very large, then the expected PN
would not be very massive, but should be H-deficient. Within this
picture the old H-rich PN would be the consequence of the first common
envelope, while the second H-deficient PN would be the consequence of
the second common envelope. The roundish shape observed in old H-rich
PNe around most H-deficient PNe might be at odds with the expected
axisymmetry of both PNe in this scenario.

Similarly in order to explain the high Ne abundances observed in the
inner ejecta of Abell 58 it was suggested that an O–Ne nova might take
place shortly after a final He shell flash
\cite{2011MNRAS.410.1870L}. In this scenario the initially more
massive star forms a ONeMg-core WD next to a AGB companion with an
intermediate separation.  During the AGB the ONeMg-core WD accretes mass
from the companion and, after the AGB, the post-AGB stars makes a PN
in a canonical single stellar evolution fashion.  Eventually, the
post-AGB star experiences a VLTP ﬂash that produces C-rich ejecta
(C/O>1 but relatively poor in Ne). The core idea behind this scenario
is that during the final flash, the new H-deficient giant starts
transferring mass to the ONeMg-core WD and this renewed accretion pushes
the WD over the limit for a nova detonation. The post-AGB star then
evolves in the usual post-VLTP sequence but surrounded by a mixture of
the ejecta from the VLTP and the Ne nova.  This Ne nova would explain
the unusual Ne composition observed in the H-deficient knots of Abell
58.  As in most binary evolution scenarios this requires a detailed
fine tuning of both the separation of the system and the accreted
masses. Moreover, under this scenario one should expect a second
eruption after the VLTP observed in 1919 (V605 Aql), which was not
observed \cite{2002Ap&SS.279..183D}.

\section{Final words}\label{sec:conclusions}
In the previous sections we have briefly reviewed the properties of
known H-deficient CSPNe and their PNe. The nature of the formation and
shaping of PNe is in general is an ongoing debate, and those PNe
with H-deficient CS are no exception. More crucially,
evolutionary scenarios for H-deficient CSPNe and their PNe face even
larger challenges, due to the diverse and exotic surface abundances
and sometimes the presence of both H-rich and H-deficient PNe with
different geometries.

 Improving our understanding of the formation of H-deficient CSPNe and PNe will require concomitant theoretical and observational efforts. On the modeling side a better understanding of non-conservative stable and unstable mass transfer events, as well as of the common envelope stage is necessary to address the viability, and frequency, of different proposed binary scenarios. Efforts to resolve the ADF issue as well as to understand the formation and evolution of dust and gas around PNe will help to improve our understanding of the effects of binarity. Existing observing facilities such as ALMA, Gemini, the GTC and the VLTI will help accomplishing these tasks. Likewise improvements in the modeling of AGB and post-AGB winds are needed to assess the frequency of late thermal pulses. This will require both improvements in the observational characterization of AGB and post-AGB winds, as well as an improvement in the hydrodynamical  modeling of the winds.

  Better, and longer, hydrodynamical simulations of the violent H-ingestion events that happen during a VLTP (and, possibly, also in white dwarf merger events) are key to improve our understanding of the born again scenario. These hydrodynamical models not only shed light on the energetics of VLTP events but are key to understand the extent of heavy element production through neutron capture processes. Quantifying the heavy-element chemical signatures  of different evolutionary scenarios might be key to disentangle the contributions of each evolutionary channel and improve the comparison with observations. Moreover, using heavy element abundances to identify possible evolutionary channels requires spectroscopic analyses in the UV. Currently the best available tool for these studies is  Cosmic Origins Spectrograph (COS) at the Hubble Space Telescope (HST). Due to the near-future dearth of appropriate instruments efforts should be done now to secure  UV spectra for the brightest representatives of the different spectral subtypes. In addition, to extract the most information from these spectra it is necessary to improve atomic data required by  state-of-the-art non-LTE model-atmosphere codes.
  
Further studies are needed to establish whether the shapes, sizes and locations of PNe with H-deficient central stars are statistically different from those with normal H-rich central stars. The large photometry and astrometric database from Gaia, together with ad-hoc catalogues \citep[e.g.][]{2020A&A...640A..10W} should help resolve this issue.

We have made an effort to present both a brief pedagogical
introduction and an updated view on the problem of the formation of
H-deficient central stars.  Regardless of our effort this review is
necessarily one seen through our own spectacles and biases. We have
tried to minimize this by providing as many references to the work of
other authors as possible, and to mark our personal positions as such
when presenting a given topic.  We hope the present review serves as a
starting point for those willing to study these fascinating objects and
their evolution.

%-----------------------------------------

%%%%%%%%%%%%%%%%%%%%%%%%%%%%%%%%%%%%%%%%%%
\vspace{6pt}

\funding{The author is partially funded by CONICET and Agencia I+D+i through grants  PIP-2971 and PICT 2020-03316 and DAAD-CONICET (2023-2024) bilateral cooperation grant RESOL-2022-2320-APN-DIR.}

\acknowledgments{M3B thanks Walter Weidmann for providing the stellar data
  presented in Fig. \ref{fig:HR-Kiel}. M3B also thanks Alejandro Córsico, Miriam Peña, Jesús Toalá, and three anonymous referees for detailed comments and suggestions over previous versions of the manuscript that helped improving the final version of the review.}

\conflictsofinterest{The author declare no conflicts of interest.} 

%%%%%%%%%%%%%%%%%%%%%%%%%%%%%%%%%%%%%%%%%%
%% Optional

%% Only for journal Encyclopedia
%\entrylink{The Link to this entry published on the encyclopedia platform.}

%\abbreviations{Abbreviations}{
%The following abbreviations are used in this manuscript:\\

%\noindent 
%\begin{tabular}{@{}ll}
%MDPI & Multidisciplinary Digital Publishing Institute\\
%DOAJ & Directory of open access journals\\
%TLA & Three letter acronym\\
%LD & Linear dichroism
%\end{tabular}
%}

%%%%%%%%%%%%%%%%%%%%%%%%%%%%%%%%%%%%%%%%%%
%% Optional
%\appendixtitles{no} % Leave argument "no" if all appendix headings stay EMPTY (then no dot is printed after "Appendix A"). If the appendix sections contain a heading then change the argument to "yes".
\appendixstart
\appendix

\section{Bona fide born again CSPNe}
\label{bonafide}
One of the strongest evidences in favour of the born again scenario for the formation of H-deficient CSPNe comes from few stars that have been detected during the late helium flash or immediately after. Due to the fast evolutionary speed during LTP and VLTP events capturing a star at the very moment they are undergoing a helium shell flash  is rare. In spite of their rarity, these stars offer a unique opportunity to validate our theories. In particular, they show that CSPNe do evolve back to a giant configuration, and they also show how stars change their spectral types and composition during this events. Due to their importance for theories of the formation of H-deficient CSPNe here we provide a brief description of their observed properties and behaviour.

\paragraph{FG Sge}

 This variable star has shown a slow and progressive rise and of its
 visual magnitude from $m_{\rm photo} = 13.6$ in 1894 to $B = 9.6$ in
 1965. In turn, the spectral type of this object changed from a B5 in
 1955 to A5 in 1967. From 1975 to date, both its luminosity as its
 color have remained approximately constant
 \citep{1998ApJS..114..133G}. These changes can be interpreted as an
 almost horizontal evolution in the HR diagram from the region of the
 CSPNe to the region of the yellow giants near the AGB. Indeed,
 \citep{1995A&A...294..453V},  infer a
 change from $\log T_{\rm eff} \simeq 4.65$ and $\log L/L_\odot \simeq
 3.4$ to $\log T_{\rm eff} \simeq 3.75$ and $\log L/L_\odot \simeq
 4.04$ in 100 years (from 1880 to 1980). The superficial abundances of
 elements related to slow neutron capture processes in this star have
 been the subject of many discussions in the literature. These results
 range from the existence of changes in approximately 2 to 3 dex
 between 1970 to 1998 \citep{1998ApJS..114..133G} to the existence of
 supersolar abundances, but constant from 1950 to the present
 \cite{2006A&A...459..885J}. During the last half century
 observational evidence that this object became H-deficient has
 accumulated.  The evolution of temperature and luminosity in the last
 100 years and the progressive decrease in surface abundance of H,
 possibly accompanied by the increase in slow neutron capture process
 elements, link FG Sge to the late thermal pulse scenario
 \cite{2006A&A...459..885J} (LTP, see Section
 \ref{Born_Again}). Since 1992 FG Sge shows events of abrupt decrease
 in luminosity, probably produced by dust condensation around the
 star, similar to those observed in RCrB stars.

 \paragraph{ V4334 Sgr}

Yukio Sakurai reported the detection of a slow eruption in Sagittarius
(V4334 Sgr) in a photograph the 20th of February of 1996. In addition,
photographs taken by Sakurai during 1993 and 1994 showed no candidate
at the same location, but that the star is visible on films taken in
January of 1995\cite{1996IAUC.6322....1N}.  Benetti et
al. \cite{1996IAUC.6325....1B} reported that both the abundances
observed in a spectrum of V4334 Sgr as well as the slow evolution of
luminosity make it very possible that V4334 Sgr is a star going
through a VLTP, which was confirmed some days later by Duerbeck and
Pollaco \cite{1996IAUC.6328....1D}, who reported the existence of an
old planetary nebula approximately 32'' in diameter in that region of
the sky. This led Duerbeck and Benetti \cite{1996ApJ...468L.111D} to
propose that, indeed, V4334 Sgr was going through a VLTP episode.
Photometric studies after its discovery showed V4334 Sgr increased its
luminosity and cooled progressively \citep{1997AJ....114.1657D},
entered a stage of successive abrupt decreases in luminosity similar
to those observed in RCrB stars \citep{2000AJ....119.2360D}, and
finally disappeared behind a cloud of dust from which it has not yet
emerged. Simultaneously, spectroscopic studies carried out in 1996
\citep{1997A&A...321L..17A,1999A&A...343..507A} showed that V4334 Sgr
was suffering changes in its surface abundances, most likely due to
the dredging of processed material coming from its inner regions.
These studies showed that during the eruption V4334 Sgr was already a
very H-deficient star, which was rich in C (both $^{12}$C and $^{13}$
C), N, and in neutron capture elements. Analysis of the
photoionization state of the nebula around V4334 Sgr by
\citep{1999A&A...344L..79K,1999MNRAS.304..127P, 2008ASPC..391..163H},
have confirmed that V4334 Sgr was, very few years ago still a hot
white dwarf star. Radio observations of V4334 Sgr carried out between
2004 and 2006 initially were interpreted as thermal emission from
freshly ionized ejecta by the central
star\cite{2005Sci...308..231H,2007A&A...471L...9V}, but were later
shown to be due to shock ionization \cite{2007A&A...471L...9V,
  2018Galax...6...79V}. The latest reanalysis all data from 2004 to
2023 suggests that V4334 had  highly variable non-thermal emission in
2004-2017, and since 2019 has started showing an increasing
contribution of thermal emission \cite{2024arXiv240718020H}. The later
might be a first hint of the photoionization of the ejecta by a
reheating central star. Due to the loss of the UV radiation from the
CS, the old PNe around the star has started to
recombine\cite{2022ApJ...939..103R}. Moreover, infrared observations
of the object in the last decade has shown the detection of new
bipolar nebulae \cite{2014ApJ...785..146H,2020ApJ...904...34H}. This
results are in line with ALMA observations that unambiguously show the
presence of an equatorial disc and bipolar outﬂows, formed in
V4334 Sgr during the 30 years after the occurrence of the
born again event \citep{2023A&A...677L...8T}. This provides important
constraints for future modeling efforts of this phenomenon as well as
the formation of axisymmetrical structures in PNe.

In short, V4334 Sgr in the last 30 years has stopped being a hot WD to cool down and increase in brightness until it became a
H-deficient yellow giant star with an RCrB type light curve, then hid
from our sight in a cloud of dust in which it might be heating up
again. This behavior is exactly what would be expected of an object
going through a VLTP event (see Section \ref{Born_Again}), so V4334
Sgr provides the strongest observational support for the born again
scenario for the formation of H-deficient stars. Moreover since the
eruption V4334 Sgr the material ejected from the star seems to have
formed both bipolar outflows and an equatorial disk inside the old
roundish PNe.

\paragraph{V605 Aql (CS of Abell 58)}
This star was discovered in 1919 and was initially classified as very
slow nova. Later observations, between 1919 and 1923, showed the star
disappearing and reappearing at irregular intervals, similar to the
observed lightcurves of RCrB stars. Based on a spectrum taken in 1921
by Lundmark \cite{1921PASP...33..314L} the star was classified as a C
star.  Modern inspection of the 1921 spectrum indicated that the
composition of the erupting star was very similar to that of an RCrB
star (with 98\% He and 1\% of C) \cite{1997AJ....114.2679C}. In
addition, many decades after the eruption, it was also recognized that
the eruption took place at the center of the old PNe Abell 58
\cite{1971ApJ...170..547F}.

Duerbeck et al. \cite{2002Ap&SS.279..183D} collected and studied
observations of the eruption of this star between 1917 and 1923,
obtaining a light curve that shows an undeniable similarity to the
light curve of V4334 Sgr—which has led to V605 Aql being described as
the “elder twin” of V4334 Sgr. Taking advantage of the fact that the
planetary nebula that surrounds this star (Abell 58) still reflects
the state of the star when it was hot enough to ionize it,
photoionization studies of the nebula \cite{2004A&A...426..145L} came
to the conclusion that the star, before its eruption, was a hot
white dwarf star. Observations carried out in 2001, and presented by \cite{2006ApJ...646L..69C}, show that V605 Aql is now a star with
a spectrum similar to that of the CSs of planetary nebula with Wolf-Rayet spectral types ([WC]-CSPN), with He/C/O abundances 54/40/5 per mass fraction). In turn, this star now shows a much more compact
He-rich nebula around it.  All of these characteristics strongly
indicate that V605 Aql is an object that experienced a VLTP event
which turned its surface composition He-, C-, and O-rich.
Consequently, V605 Aql offers a strong validation of the VLTP scenario
as a channel for the formation of H-deficient CSPNe.  Moreover, given
the similar He-dominated composition shown by both V605 Aql and V4334
Sgr at the moment of the eruption
\cite{1997AJ....114.2679C,1997A&A...321L..17A,1999A&A...343..507A}
V605 Aql suggests that V4334 Sgr, whose composition at the moment of
the outburst did not match that of the intershell abundances AGB
stars, might soon become He-, C-, and O-rich.  As in the
case of V4334 Sgr, studies of the inner ejecta in the IR
\cite{2008A&A...479..817H} and with ALMA \cite{2022ApJ...925L...4T}
indicate the ejection of bipolar outflows, and the formation of
axysimetric structures after the eruption.
 Multiepoch optical spectroscopic observations of the H-deﬁcient ejecta around V605 Aql shows that the ejecta  
  has experienced significant  brightening from 1996 to 2021. This changes have been attributed to shocks in the bipolar H-deficient outﬂow \cite{2022ApJ...934...18M}. Integral field spectroscopic observations  obtained on 2022 using MEGARA at the GTC, in combination with ALMA and Hubble Space Telescope data, lead \cite{2024A&A...684A.107M} to propose  very complex structure around V605 Aql. 
The structure consist of an hourglass structure for the ionized material,
around a high-velocity polar components together with an expanding toroidal molecular and dusty structure. Such complex structure has been interpreted by \cite{2024A&A...684A.107M}  as a proof of the presence of an unseen close stellar or sub-stellar companion.

\paragraph{Hen 3-1357, SAO 244567 (CS of the Stingray PN)}

First reported in 1970 as a B type star \cite{1970MmRAS..73..153W},
Hen 3-1357 was recognized in 1993 as a very fast evolving star
\cite{1993A&A...267L..19P}. Based on the optical spectrum, and its UBV
colours, it was concluded that Hen 3-1357 had an effective temperature
of $T_{\rm eff} \simeq 21$ kK in 1971
\cite{1993A&A...267L..19P,1995A&A...300L..25P}. Optical and
ultraviolet spectra between 1990 and 1992 revealed that Hen 3-1357 had
ionized its surrounding nebula within only two decades (the Stingray
Nebula) \cite{1993A&A...267L..19P,1995A&A...300L..25P}.  Reindl et
al. \cite{2014A&A...565A..40R}, presented first quantitative spectral
analyses of all available spectra and found that the CS had increased
its $T_{\rm eff}$ from 38000 K in 1988 to a peak value of 60000 K in
2002. During this period its surface gravity increased continuously
from $\log g = 4.8$ to $\log g =6$, which implies a significant
drop in luminosity. In a follow up study, these authors
\cite{2017MNRAS.464L..51R} found that Hen 3-1357, after the initial
rapid heating and contraction, had cooled and expanded significantly
from 2002 to 2006, reaching values of $T_{\rm eff}=50000$ K and $\log
g =5.5$ in 2006. These observations confirmed the suspicion that Hen
3-1357 is undergoing a LTP, as such change in the
evolution of the effective temperature and gravity can only be
explained within the LTP scenario.  The sudden change in the evolution
of the star due to the late He flash is also notable in the lightcurve
of the star , which shows a steady decline from 1889 to 1980, followed
by a sudden fast fading from 1980 to 2006 \cite{2015ApJ...812..133S}.
The evolutionary speed suggested a CS mass between 0.53 and 0.56
$M_\odot$ \cite{2017MNRAS.464L..51R}. However, none of the current LTP
models can fully reproduce the evolution of all characteritics of the
star \cite{2017MNRAS.464L..51R,2021MNRAS.504..667L}. The comparison of
the nebular abundances with those predicted by stellar evolution and
nucleosynthesis models at the end of the AGB phase suggest that the CS
had an initial mass lower than 1.5 $M_\odot$.  The sudden cooling of
the CS was followed by large decreases in its nebular emission-line
ﬂuxes. It has been now confirmed that the nebula is recombining due to
the dissapearance of the ionizing source \citep{2022MNRAS.515.1459P,
  2021ApJ...907..104B}.  From the analysis of the physical conditions
of the nebulae some authors concluded that the effective temperature
of Hen 3-1357 has decreased from about 60000 K in 2002 to less than 40000
K in 2021\citep{2022MNRAS.515.1459P}.

Based on our understading of the scenario we expect Hen 3-1357 to
continue cooling for some decades, reaching a giant configuration
again. Whether this star will become H-deficient or not in the future
depends on whether third dredge up events take place once the star is
back on the AGB, something that depends strongly on the mass of the
CS and on the physics of CBM which is
not well understood at the moment (see Section \ref{Born_Again}).

\paragraph{Other stars discussed in connection with the born again
  scenario} In addition to the objects mentioned above, other
stars/eruptions have been mentioned in connection with the born again
scenario. Among them V838 Monocerotis, CK Vul, and NSV43434 have been
discused as possible born again
events\citep{2002MNRAS.332L..35E,2005MNRAS.361..695L,2011ApJ...743L..33M},
but were later shown to be novae, stellar mergers or other types of
stellar
transients\cite{2005A&A...441.1099T,2006A&A...451..223T,2012PASP..124.1262B,2024A&A...685A..49T,2019A&A...630A..75P}. Moreover,
the central star HuBi 1 has been identified by some studies as
evolving fast through the HR diagram after a VLTP event
\cite{2000A&A...362.1008G,2001A&A...367..983P,2005RMxAA..41..423P,2023ApJ...955..151R},
although this interpreation is still under debate and alternative
interpretations have been suggested
\cite{2018NatAs...2..784G,2021MNRAS.505.3883T,2022MNRAS.512.4003M}.

%\section[\appendixname~\thesection]{}
%\subsection[\appendixname~\thesubsection]{}
%The appendix is an optional section that can contain details and data supplemental to the main text---for example, explanations of experimental details that would disrupt the flow of the main text but nonetheless remain crucial to understanding and reproducing the research shown; figures of replicates for experiments of which representative data are shown in the main text can be added here if brief, or as Supplementary Data. Mathematical proofs of results not central to the paper can be added as an appendix.

%\begin{table}[H] 
%\caption{This is a table caption.\label{tab5}}
%\newcolumntype{C}{>{\centering\arraybackslash}X}
%\begin{tabularx}{\textwidth}{CCC}
%\toprule
%\textbf{Title 1}	& \textbf{Title 2}	& \textbf{Title 3}\\
%\midrule
%Entry 1		& Data			& Data\\
%Entry 2		& Data			& Data\\
%\bottomrule
%\end{tabularx}
%\end{table}

%\section[\appendixname~\thesection]{}
%All appendix sections must be cited in the main text. In the appendices, Figures, Tables, etc. should be labeled, starting with ``A''---e.g., Figure A1, Figure A2, etc.

%%%%%%%%%%%%%%%%%%%%%%%%%%%%%%%%%%%%%%%%%%
\begin{adjustwidth}{-\extralength}{0cm}

\reftitle{References}

\PublishersNote{}
\end{adjustwidth}
\end{document}